\documentclass[10pt,preprint]{aastex}

\newcommand{\binomial}{\left( {\begin{array}{*{20}c} N \\ n \\ \end{array}} \right)}

\begin{document}
  
  \title{Some Aspects of Measurement Error in Linear Regression 
  of Astronomical Data}

  \author{Brandon C. Kelly}
  \email{bkelly@as.arizona.edu}
  \affil{Steward Observatory, University of Arizona, 933 N Cherry Ave,
  Tucson, AZ 85721}
  
  \begin{abstract}

    I describe a Bayesian method to account for measurement errors in
    linear regression of astronomical data. The method allows for
    heteroscedastic and possibly correlated measurement errors, and
    intrinsic scatter in the regression relationship. The method is
    based on deriving a likelihood function for the measured data, and
    I focus on the case when the intrinsic distribution of the
    independent variables can be approximated using a mixture of
    Gaussians. I generalize the method to incorporate multiple
    independent variables, non-detections, and selection effects
    (e.g., Malmquist bias). A Gibbs sampler is described for
    simulating random draws from the probability distribution of the
    parameters, given the observed data. I use simulation to compare
    the method with other common estimators. The simulations
    illustrate that the Gaussian mixture model outperforms other
    common estimators and can effectively give constraints on the
    regression parameters, even when the measurement errors dominate
    the observed scatter, source detection fraction is low, or the
    intrinsic distribution of the independent variables is not a
    mixture of Gaussians. I conclude by using this method to fit the
    X-ray spectral slope as a function of Eddington ratio using a
    sample of 39 $z \lesssim 0.8$ radio-quiet quasars. I confirm the
    correlation seen by other authors between the radio-quiet quasar
    X-ray spectral slope and the Eddington ratio, where the X-ray
    spectral slope softens as the Eddington ratio increases. IDL
    routines are made available for performing the regression.

  \end{abstract}
  
  
  \section{INTRODUCTION}

  \label{s-intro}

  Linear regression is one of the most common statistical techniques
  used in astronomical data analysis. In general, linear regression in
  astronomy is characterized by intrinsic scatter about the regression
  line, and measurement errors in both the independent and dependent
  variables. The source of intrinsic scatter is variations in the
  physical properties of astronomical sources that are not completely
  captured by the variables included in the regression. It is
  important to correctly account for both measurement error and
  intrinsic scatter, as both aspects can have a non-negligible effect
  on the regression results. In particular, ignoring the intrinsic
  scatter and weighting the data points solely by the measurement
  errors can result in the higher-precision measurements being given
  disproportionate influence on the regression results. Furthermore,
  when the independent variable is measured with error, the ordinary
  least squares (OLS) estimate of the regression slope is biased
  toward zero \citep[e.g.,][]{fuller87,bces,fox97}. When there are
  multiple independent variables, measurement error can have an even
  stronger and more unpredictable effect \citep{fox97}. In addition,
  the existence of non-detections, referred to as `censored data', in
  the data set will result in additional complications
  \citep[e.g.,][]{isobe86}. Therefore, when performing regression, it
  is essential to correctly account for the measurement errors and
  intrinsic scatter in order to ensure that the data analysis, and
  thus the scientific conclusions based on it, are trustworthy.

  Many methods have been proposed for performing linear regression
  when intrinsic scatter is present and both variables are measured
  with error. These include methods that correct the observed moments
  of the data \citep[e.g.,][]{fuller87,bces,freed04}, minimize an
  `effective' $\chi^2$ statistic
  \citep[e.g.,][]{clut67,bark74,numrec,trem02}, assume a probability
  distribution for the true independent variable values
  \citep[so-called `structural equation models',
  e.g.,][]{schafer87,schafer01,roy06}; Bayesian approaches to these
  models have also been developed
  \citep[e.g.,][]{zell71,gull89,dell95,carroll99,scheines99}. In
  addition, methods have been proposed to account for measurement
  error in censored regression \citep[e.g.,][]{stap84,weiss93}. The
  most commonly used methods in astronomy are the BCES estimator
  \citep{bces} and the `FITEXY' estimator \citep{numrec}. Both methods
  have their advantages and disadvantages, some of which have been
  pointed out by \citet{trem02}. However, neither method is applicable
  when the data contain non-detections.

  In this work I describe a Bayesian method for handling measurement
  errors in astronomical data analysis. My approach starts by
  computing the likelihood function of the complete data, i.e., the
  likelihood function of both the unobserved true values of the data
  and the measured values of the data. The measured data likelihood is
  then found by integrating the likelihood function for the complete
  data over the unobserved true values
  \citep[e.g.,][]{lit02,gelman04}. This approach is known as
  `structural equation modelling' of measurement error problems, and
  has been studied from both a frequentist approach
  \citep[e.g.,][]{fuller87,carroll95,schafer01,aitken02} and a
  Bayesian approach \citep[e.g.,][]{muller97,rich97,rich02}. In this
  work, I extend the statistical model of \citet{carroll99} to allow
  for measurement errors of different magnitudes (i.e.,
  `heteroscedastic' errors), non-detections, and selection effects, so
  long as the selection function can be modelled mathematically. Our
  method models the distribution of independent variables as a
  weighted sum of Gaussians. The mixture of Gaussians model allows
  flexibility when estimating the distribution of the true values of
  the independent variable, thus increasing its robustness against
  model mispecification \citep[e.g.,][]{huang06}. The basic idea is
  that one can use a suitably large enough number of Gaussians to
  accurately approximate the true distribution of independent
  variables, even though in general the individual Gaussians have no
  physical meaning.

  The paper is organized as follows. In \S~\ref{s-notation} we
  summarize some notation, and in \S~\ref{s-measerr} I review the
  effects of measurement error on the estimates for the regression
  slope and correlation coefficient. In \S~\ref{s-statmod} I describe
  the statistical model and derive the likelihood functions, and in
  \S~\ref{s-dcissues} I describe how to incorporate knowledge of the
  selection effects and account for non-detections. In
  \S~\ref{s-prior} I describe the prior distribution for this model,
  and in \S~\ref{s-markov} I describe a Gibbs sampler for sampling
  from the posterior distributions. In \S~\ref{s-sims} I use
  simulation to illustrate the effectiveness of this structural model
  and compare with the OLS, BCES($Y|X$), and FITEXY
  estimators. Finally, in \S~\ref{s-data} I illustrate the method
  using astronomical data by performing a regression of the X-ray
  photon index, $\Gamma_X$, on the Eddington ratio using a sample of
  39 $z < 0.83$ radio-quiet quasars. Sections \ref{s-statmod},
  \ref{s-dcissues}, and \ref{s-compmeth} are somewhat technical, and
  the reader who is uninterested in the mathematical and computational
  details may skip to them.

 \section{NOTATION}

  \label{s-notation}

  I will use the common statistical notation that an estimate of a
  quantity is denoted by placing a `hat' above it; e.g.,
  $\hat{\theta}$ is an estimate of the true value of the parameter
  $\theta$. In general, greek letters will denote the true value of a
  quantity, while roman letters will denote the contaminated measured
  value. I will frequently refer to the `bias' of an estimator. The
  bias of an estimator is $E(\hat{\theta}) - \theta_0$, where
  $E(\hat{\theta})$ is the expectation value of the estimator
  $\hat{\theta}$, and $\theta_0$ is the true value of $\theta$. An
  unbiased estimator is one such that $E(\hat{\theta}) = \theta_0$. 

  I will denote a normal density with mean $\mu$ and variance
  $\sigma^2$ as $N(\mu, \sigma^2)$, and I will denote as $N_p (\mu,
  \Sigma)$ a multivariate normal density with $p$-element mean vector
  $\mu$ and $p \times p$ covariance matrix $\Sigma$. If I want to
  explicitly identify the argument of the Gaussian function, I will
  use the notation $N(x|\mu,\sigma^2)$, which should be understood to
  be a Gaussian with mean $\mu$ and variance $\sigma^2$ as a function
  of $x$. Following \citet{gelman04}, I denote the scaled
  inverse-$\chi^2$ density as Inv-$\chi^2(\nu, s^2)$, where $\nu$ is
  the degrees of freedom and $s^2$ is the scale parameter, and we
  denote the inverse-Wishart as Inv-${\rm Wishart}_{\nu}(S)$, where
  $\nu$ is the degrees of freedom and $S$ is the scale matrix. The
  inverse-Wishart distribution can be thought of as a multivariate
  generalization of the scaled inverse-$\chi^2$ distribution. I will
  often use the common statistical notation where ``$\sim$'' means
  ``is drawn from'' or ``is distributed as''. For example, $x \sim
  N(\mu, \sigma^2)$ states that $x$ is drawn from a normal density
  with mean $\mu$ and variance $\sigma^2$.

  \section{EFFECT OF MEASUREMENT ERROR ON CORRELATION AND REGRESSION}

  \label{s-measerr}

  It is well known that measurement error can attenuate the estimate
  of the regression slope and correlation coefficient
  \citep[e.g.,][]{fuller87,fox97}. For completeness, I give a brief
  review of the effect of measurement error on correlation and
  regression analysis for the case of one independent variable.

  Denote the independent variable as $\xi$ and the dependent variable
  as $\eta$; $\xi$ and $\eta$ are also referred to as the `covariate'
  and the `response', respectively. I assume that $\xi$ is a random
  vector of $n$ data points drawn from some probability
  distribution. The dependent variable, $\eta$, depends on $\xi$
  according to the usual additive model:
  \begin{equation}
    \eta_i = \alpha + \beta \xi_i + \epsilon_i
    \label{eq-adderr}
  \end{equation}
  Here, $\epsilon_i$ is a random variable representing the intrinsic
  scatter in $\eta_i$ about the regression relationship, and $(\alpha,
  \beta)$ are the regression coefficients. The mean of $\epsilon$ is
  assumed to be zero, and the variance of $\epsilon$ is assumed to be
  constant and is denoted as $\sigma^2$. We do not observe the actual
  values of $(\xi, \eta)$, but instead observe values $(x, y)$ which
  are measured with error. The measured values are assumed to be
  related to the actual values as
  \begin{eqnarray}
    x_i & = & \xi_i + \epsilon_{x,i} \label{eq-xerr} \\
    y_i & = & \eta_i + \epsilon_{y,i} \label{eq-yerr},
  \end{eqnarray}
  where $\epsilon_{x,i}$ and $\epsilon_{y,i}$ are the random
  measurement errors on $x_i$ and $y_i$, respectively. In general, the
  errors are normally distributed with known variances
  $\sigma_{x,i}^2$ and $\sigma^2_{y,i}$, and covariance
  $\sigma_{xy,i}$. For simplicity, throughout the rest of this section
  I assume that $\sigma^2_x,\sigma^2_y,$ and $\sigma_{xy}$ are the
  same for each data point.

  When the data are measured without error, the least-squares estimate
  of the regression slope, $\hat{\beta}_{OLS}$, and the estimated
  correlation coefficient, $\hat{\rho}$, are
  \begin{eqnarray}
    \hat{\beta}_{OLS} & = & \frac{Cov(\xi,\eta)}{Var(\xi)} \label{eq-betaols} \\
    \hat{\rho}        & = & \frac{Cov(\xi,\eta)}{\sqrt{Var(\xi) Var(\eta)}} = 
    \hat{\beta}_{OLS} \sqrt{\frac{Var(\xi)}{Var(\eta)}}. \label{eq-rho}
  \end{eqnarray}
  Here, $Cov(\xi,\eta)$ is the sample covariance between $\xi$ and
  $\eta$, and $Var(\xi)$ is the sample variance of $\xi$. When the
  data are measured with error, the least-squares estimate of the
  regression slope, $\hat{b}_{OLS}$, and the estimated correlation
  coefficient, $\hat{r}$, become
  \begin{eqnarray}
    \hat{b}_{OLS} & = & \frac{Cov(x,y)}{Var(x)} = 
    \frac{Cov(\xi,\eta) + \sigma_{xy}}{Var(\xi) + \sigma^2_x} \label{eq-bols} \\
    \hat{r}       & = & \frac{Cov(x,y)}{\sqrt{Var(x) Var(y)}} = 
    \frac{Cov(\xi,\eta) + \sigma_{xy}}{\sqrt{(Var(\xi) + \sigma^2_x)(Var(\eta) + \sigma^2_y)}}. 
    \label{eq-rhat}
  \end{eqnarray}
  From these equations it is apparent that the estimated slope and
  correlation are biased when the data are measured with error.

  It is informative to assess the effect of measurement error in terms
  of the ratios $R_x = \sigma_x^2 / Var(x), R_y = \sigma^2_y / Var(y),
  R_{xy} = \sigma_{xy} / Cov(x,y)$, as these quantities can be
  calculated from the data. The fractional bias in the estimated slope
  and correlation may then be expressed as
  \begin{eqnarray}
    \frac{\hat{b}}{\hat{\beta}} & = & \frac{1 - R_x}{1 - R_{xy}} \label{eq-frac_bbias} \\
    \frac{\hat{r}}{\hat{\rho}}  & = & \frac{\sqrt{(1 - R_x)(1 - R_y)}}{1 - R_{xy}}. 
    \label{eq-frac_rbias}
  \end{eqnarray}
  From Equations (\ref{eq-frac_bbias}) and (\ref{eq-frac_rbias}) it is
  apparent that measurement errors have the following effects. First,
  covariate measurement error reduces the magnitude of the observed
  correlation between the independent variable and the response, as
  well as biasing the estimate of the slope towards zero. Second,
  measurement error in the response also reduces the magnitude of the
  observed correlation between the variables. Third, if the
  measurement errors are correlated the effects depend on the sign of
  this correlation. If the measurement error correlation has the same
  sign as the intrinsic correlation between $\xi$ and $\eta$, then the
  measurement errors cause a spurious increase in the observed
  correlation; otherwise the measurement errors cause a spurious
  decrease in the observed correlation. The magnitude of these effects
  depend on how large the measurement errors are compared to the
  observed variance in $x$ and $y$.

\begin{figure}
    \begin{center}
      \scalebox{0.5}{\rotatebox{90}{\plotone{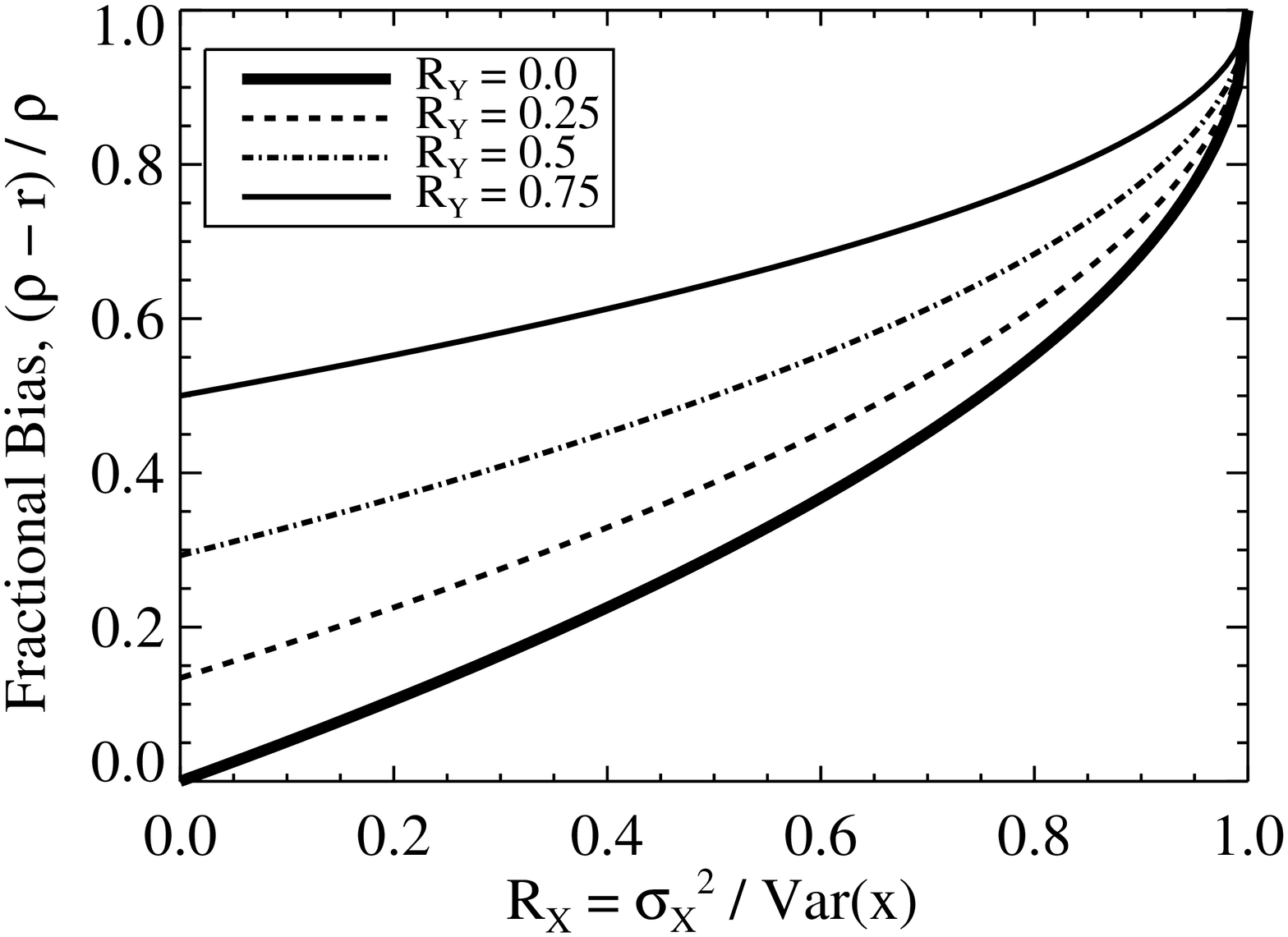}}}
      \caption{The fractional bias in the correlation coefficient when
      the data are contaminated with measurement error. The fractional
      bias is shown as a function of the contribution of measurement
      error to the observed variance in both $x$ and $y$, for
      uncorrelated measurement errors. When the measurement errors
      make up $\sim 50\%$ of the observed variance in both $x$ and
      $y$, the observed correlation coefficient is reduced by about
      $\sim 50\%$. \label{f-bias}}
    \end{center}
  \end{figure}

  In Figure \ref{f-bias} I plot the fractional bias in the
  correlation coefficient, $(\hat{\rho} - \hat{r}) / \hat{\rho}$, as a
  function of $R_x$ and $R_y$ when the errors are uncorrelated. As can
  be seen, measurement error can have a significant effect on the
  estimation of the linear correlation coefficient. For example, when
  $R_x \approx 0.5$ and $R_y \approx 0.5$, the estimated correlation
  is $\approx 50$\% lower than the true correlation. Therefore,
  interpretation of correlation coefficients and regression slopes
  must be approached with caution when the data have been contaminated
  by measurement error. To ensure accurate results, it is necessary to
  employ statistical methods that correct for the measurement errors.

  \section{THE STATISTICAL MODEL}

  \label{s-statmod}

  \subsection{Regression with One Independent Variable}

  \label{s-normreg}

  I assume that the independent variable, $\xi$, is drawn from a
  probability distribution $p(\xi|\psi)$, where $\psi$ denotes the
  parameters for this distribution.  The dependent variable is then
  drawn from the conditional distribution of $\eta$ given $\xi$,
  denoted as $p(\eta|\xi,\theta)$; $\theta$ denotes the parameters for
  this distribution. The joint distribution of $\xi$ and $\eta$ is
  then $p(\xi,\eta|\psi,\theta) = p(\eta|\xi,\theta) p(\xi|\psi)$. In
  this work I assume the normal linear regression model given by
  Equation (\ref{eq-adderr}), and thus $p(\eta|\xi,\theta)$ is a
  normal density with mean $\alpha + \beta \xi$ and variance
  $\sigma^2$, and $\theta = (\alpha, \beta, \sigma^2)$.

  Since the data are a randomly observed sample, we can derive the
  likelihood function for the measured data. The likelihood function
  of the measured data, $p(x,y|\theta,\psi)$, is obtained by
  integrating the complete data likelihood over the missing data,
  $\xi$ and $\eta$ \citep[e.g.,][]{lit02,gelman04}:
  \begin{equation}
    p(x,y|\theta, \psi) = \int \int p(x,y,\xi,\eta|\theta, \psi)\ d\xi\ d\eta.
    \label{eq-obslik1}
  \end{equation}
  Here, $p(x,y,\xi,\eta|\theta, \psi)$ is the complete data likelihood
  function. Because of the hierarchical structure inherent in the measurement
  error model, it is helpful to decompose the complete data likelihood
  into conditional probability densities:
  \begin{equation}
    p(x,y|\theta, \psi) = \int \int p(x,y|\xi, \eta) p(\eta|\xi, \theta)
    p(\xi|\psi)\ d\xi\ d\eta.
    \label{eq-obslik2}
  \end{equation}
  The density $p(x,y|\xi,\eta)$ describes the joint distribution of
  the measured values $x$ and $y$ at a given $\xi$ and $\eta$, and
  depends on the assumed distribution of the measurement errors,
  $\epsilon_x$ and $\epsilon_y$. In this work I assume Gaussian
  measurement error, and thus $p(x_i,y_i|\xi_i,\eta_i)$ is a
  multivariate normal density with mean $(\xi_i,\eta_i)$ and
  covariance matrix $\Sigma_i$, where $\Sigma_{11,i} = \sigma^2_{y,i},
  \Sigma_{22,i} = \sigma^2_{x,i},$ and $\Sigma_{12,i} =
  \sigma_{xy,i}$.  The statistical model may then be conveniently expressed
  hierarchically as
  \begin{eqnarray}
    \xi_i & \sim & p(\xi|\psi) \label{eq-hierxi} \\
    \eta_i|\xi_i & \sim & N(\alpha + \beta \xi_i, \sigma^2) \label{eq-hiereta} \\
    y_i,x_i|\eta_i,\xi_i & \sim & N_2([\eta_i,\xi_i],\Sigma_i) \label{eq-hierxy}
  \end{eqnarray}
  Note that if $x_i$ is measured without error, then $p(x_i|\xi_i)$ is
  a Dirac delta function, and $p(x_i,y_i|\xi_i,\eta_i) = p(y_i|\eta_i)
  \delta(x_i - \xi_i)$. An equivalent result holds if $y_i$ is
  measured without error.

  Equation (\ref{eq-obslik2}) may be used to obtain the observed data
  likelihood function for any assumed distribution of $\xi$. In this
  work, I model $p(\xi|\psi)$ as a mixture of $K$ Gaussians,
  \begin{equation}
    p(\xi_i|\psi) = \sum_{k=1}^{K} \frac{\pi_k}{\sqrt{2\pi \tau^2_k}} 
    \exp \left\{ -\frac{1}{2} \frac{(\xi_i - \mu_k)^2}{\tau^2_k} \right \},
    \label{eq-xidist}
  \end{equation}
  where $\sum_{k=1}^{K} \pi_k = 1$. Note that, $\pi_k$ may be
  interpreted as the probability of drawing a data point from the
  $k^{\rm th}$ Gaussian. I will use the convenient notation $\pi =
  (\pi_1,\ldots,\pi_K), \mu = (\mu_1,\ldots,\mu_K),$ and $\tau^2 =
  (\tau^2_1,\ldots,\tau^2_K)$; note that $\psi = (\pi,\mu,\tau^2)$. It
  is useful to model $p(\xi|\psi)$ using this form because it is
  flexible enough to adapt to a wide variety of distributions, but is
  also conjugate for the regression relationship
  (Eq.[\ref{eq-adderr}]) and the measurement error distribution, thus
  simplifying the mathematics.

  Assuming the Gaussian mixture model for $p(\xi|\psi)$, the measured
  data likelihood for the $i^{\rm th}$ data point can be directly
  calculated using Equation (\ref{eq-obslik2}). Denoting the measured
  data as ${\bf z} = (y, x)$, the measured data likelihood function
  for the $i^{\rm th}$ data point is then a mixture of bivariate
  normal distributions with weights $\pi$, means $\zeta =
  (\zeta_1,\ldots,\zeta_K)$, and covariance matrices $V_i =
  (V_{1,i},\ldots,V_{K,i})$. Because the data points are statistically
  independent, the full measured data likelihood is then the product
  of the likelihood functions for the individual data points:
  \begin{eqnarray}
    p(x,y|\theta,\psi) & = & \prod_{i=1}^{n} \sum_{k=1}^{K} \frac{\pi_k}{2\pi |V_{k,i}|^{1/2}}
    \exp \left\{ -\frac{1}{2} ({\bf z}_i - \zeta_k)^T V_{k,i}^{-1} ({\bf z}_i - \zeta_k)
      \right\} \label{eq-bobslik} \\
    \zeta_k & = & (\alpha + \beta \mu_k, \mu_k) \label{eq-zeta} \\
    V_{k,i} & = & \left( \begin{array}{cc}
      \beta^2 \tau_k^2 + \sigma^2 + \sigma^2_{y,i} & \beta \tau_k^2 + \sigma_{xy,i} \\
      \beta \tau_k^2 + \sigma_{xy,i} & \tau_k^2 + \sigma^2_{x,i} \end{array} \right).
    \label{eq-sigmaz}
  \end{eqnarray}
  Here, ${\bf z}^T$ denotes the transpose of ${\bf z}$. Equation
  (\ref{eq-bobslik}) may be maximized to compute the
  maximum-likelihood estimate (MLE). When $K > 1$, the
  expectation-maximization (EM) algorithm \citep{em} is probably the
  most efficient tool for calculating the MLE. \citet{roy06} describe
  an EM algorithm when $p(\xi)$ is assumed to be a mixture of normals
  and the measurement error distribution is multivariate $t$, and
  their results can be extended to the statistical model described in
  this work.

  It is informative to decompose the measured data likelihood,
  $p(x_i,y_i|\theta,\psi) = p(y_i|x_i,\theta,\psi) p(x_i|\psi)$, as
  this representation is useful when the data contain non-detections
  (cf., \S~\ref{s-nondet}). The marginal distribution of $x_i$ is
  \begin{equation}
    p(x_i|\psi) = \sum_{k=1}^{K} \frac{\pi_k}{\sqrt{2\pi (\tau^2_k + \sigma^2_{x,i})}}
    \exp \left\{ -\frac{1}{2} \frac{(x_i - \mu_k)^2}{\tau^2_k + \sigma^2_{x,i}} \right\},
    \label{eq-xmarginal}
  \end{equation}
  and the conditional distribution of $y_i$ given $x_i$ is
  \begin{eqnarray}
    p(y_i|x_i,\theta,\psi) & = & \sum_{k=1}^K \frac{\gamma_k}{\sqrt{2\pi Var(y_i|x_i,k)}}
    \exp \left\{ -\frac{1}{2} \frac{[y_i - E(y_i|x_i,k)]^2}{Var(y_i|x_i,k)}
      \right\} \label{eq-cdensyx} \\
    \gamma_k & = & \frac{\pi_k N(x_i|\mu_k, \tau^2_k + \sigma^2_{x,i})}
	  {\sum_{j=1}^K \pi_j N(x_i|\mu_j, \tau^2_j + \sigma^2_{x,i})}
	  \label{eq-cdenskx} \\
    E(y_i|x_i,k) & = &  \alpha + \left( \frac{\beta \tau^2_k + \sigma_{xy,i}}
    {\tau^2_k + \sigma^2_{x,i}} \right) x_i + \left( \frac{\beta \sigma^2_{x,i} - \sigma_{xy,i}}
    {\tau_k^2 + \sigma^2_{x,i}} \right) \mu_k \label{eq-cmeanyx} \\
    Var(y_i|x_i,k) & = & \beta^2 \tau^2_k + \sigma^2 + \sigma^2_{y,i} - 
    \frac{(\beta \tau^2_k - \sigma_{xy,i})^2}{\tau^2_k + \sigma^2_{x,i}} \label{eq-cvary}.
  \end{eqnarray}
  Here, $\gamma_k$ can be interpreted as the probability that the
  $i^{\rm th}$ data point was drawn from the $k^{\rm th}$ Gaussian
  given $x_i$, $E(y_i|x_i,k)$ gives the expectation value of $y_i$ at
  $x_i$, given that the data point was drawn from the $k^{\rm th}$
  Gaussian, and $Var(y_i|x_i,k)$ gives the variance in $y_i$ at $x_i$,
  given that the data point was drawn from the $k^{\rm th}$ Gaussian.

  \subsection{Relationship between Uniformly Distributed Covariates and 
    Effective $\chi^2$ Estimators}

  \label{s-unifreg}

  It is informative to investigate the case where the distribution of
  $\xi$ is assumed to be uniform, $p(\xi) \propto 1$. Interpreting
  $p(\xi)$ as a `prior' on $\xi$, one may be tempted to consider
  assuming $p(\xi) \propto 1$ as a more objective alternative to the
  normal distribution. A uniform distribution for $\xi$ may be
  obtained as the limit $\tau^2 \rightarrow \infty$, and thus the
  likelihood function for $p(\xi) \propto 1$ can be calculated from
  Equation (\ref{eq-cdensyx}) by taking $\tau^2 \rightarrow \infty$
  and $K = 1$. When the measurement errors are uncorrelated, the
  likelihood for uniform $p(\xi)$ is
  \begin{equation}
    p(y|x,\theta) = \prod_{i=1}^{n}\frac{1}{\sqrt{2 \pi (\sigma^2 + \sigma^2_{y,i} + 
	\beta^2 \sigma^2_{x,i})}} \exp \left\{-\frac{1}{2}\frac{(y_i - \alpha - \beta x_i)^2}
	{\sigma^2 + \sigma^2_{y,i} + \beta^2 \sigma^2_{x,i}} \right\} .
    \label{eq-uniflik}
  \end{equation}
  The argument of the exponential is the FITEXY goodness of fit
  statistic, $\chi^2_{EXY}$, as modified by \citet{trem02} to account
  for intrinsic scatter; this fact has also been recognized by
  \citet{weiner06}. Despite this connection, minimizing $\chi^2_{EXY}$
  is not the same as maximizing the conditional likelihood of $y$
  given $x$, as both $\beta$ and $\sigma^2$ appear in the
  normalization of the likelihood function as well.

  For a given value of $\sigma^2$, minimizing $\chi^2_{EXY}$ can be
  interpreted as minimizing a weighted sum of squared errors, where
  the weights are given by the variances in $y_i$ at a given $x_i$,
  and one assumes a uniform distribution for $\xi$. Unfortunately,
  this is only valid for a fixed value of $\sigma^2$. Moreover, little
  is known about the statistical properties of the FITEXY estimator,
  such as its bias and variance, although bootstrapping
  \citep[e.g.,][]{efron79,dav97} may be used to estimate
  them. Furthermore, it is ambiguous how to calculate the FITEXY
  estimates when there is an intrinsic scatter term. The FITEXY
  goodness-of-fit statistic, $\chi^2_{EXY}$, cannot be simultaneously
  minimized with respect to $\alpha,\beta,$ and $\sigma^2$, as
  $\chi^2_{EXY}$ is a strictly decreasing function of $\sigma^2$. As
  such, it is unclear how to proceed in the optimization beyond an
  \emph{ad hoc} approach. Many authors have followed the approach
  adopted by \citet{trem02} and increase $\sigma^2$ until
  $\chi^2_{EXY} / (n - 2) = 1$, or assume $\sigma^2 = 0$ if
  $\chi^2_{EXY} / (n - 2) < 1$.

  Despite the fact that minimizing $\chi_{EXY}^2$ is not the same as
  maximizing Equation (\ref{eq-uniflik}), one may still be tempted to
  calculate a MLE based on Equation (\ref{eq-uniflik}). However, it
  can be shown that if one assumes $p(\xi) \propto 1$, and if all of
  the $x$ and $y$ have the same respective measurement error
  variances, $\sigma^2_x$ and $\sigma^2_y$, the MLE estimates for
  $\alpha$ and $\beta$ are just the ordinary least squares estimates
  \citep{zell71}. While this is not necessarily true when the
  magnitudes of the measurement errors vary between data points, one
  might expect that the MLE will behave similarly to the OLS
  estimate. I confirm this fact using simulation in
  \S~\ref{s-sims1}. Unfortunately, this implies that the MLE for
  $p(\xi) \propto 1$ inherits the bias in the OLS estimate, and thus
  nothing is gained. Furthermore, as argued by \citet{gull89}, one can
  easily be convinced that assuming $p(\xi) \propto 1$ is incorrect by
  examining a histogram of $x$.

  \subsection{Regression with Multiple Independent Variables}
  
  \label{s-multreg}

  The formalism developed in \S~\ref{s-normreg} can easily be
  generalized to multiple independent variables. In this case Equation
  (\ref{eq-adderr}) becomes
  \begin{equation}
    \eta_i = \alpha + \beta^T \xi_i + \epsilon_i,
    \label{eq-madderr}
  \end{equation}
  where $\beta$ is now a $p$-element vector and $\xi_i$ is a
  $p$-element vector containing the values of the independent
  variables for the $i^{\rm th}$ data point. Similar to before, we
  assume that the distribution of $\xi_i$ can be approximated using a
  mixture of $K$ multivariate normal densities with $p$-element mean
  vectors $\mu = (\mu_1,\ldots,\mu_K)$, $p \times p$ covariance
  matrices $T = (T_1,\ldots,T_K)$, and weights $\pi =
  (\pi_1,\ldots,\pi_K)$. The measured value of $\xi_i$ is the
  $p$-element vector ${\bf x}_i$, and the Gaussian measurement errors
  on $(y_i,{\bf x}_i)$ have $(p + 1) \times (p + 1)$ covariance matrix
  $\Sigma_i$. The statistical model is then
  \begin{eqnarray}
    \xi_i & \sim & \sum_{k=1}^{K} \pi_k N_p(\mu_k, T_k) \label{eq-mhierxi} \\
    \eta_i|\xi_i & \sim & N(\alpha + \beta^T \xi_i, \sigma^2) \label{eq-mhiereta} \\
    y_i,{\bf x}_i|\eta_i,\xi_i & \sim & N_{p+1}([\eta_i,\xi_i], \Sigma_i). \label{eq-mhierxy} \\
  \end{eqnarray}

  Denoting ${\bf z}_i = (y_i, {\bf x}_i)$, the measured data
  likelihood is
  \begin{eqnarray}
    p(x,y|\theta,\psi) & = & \prod_{i=1}^{n} \sum_{k=1}^K \frac{\pi_k}{(2\pi)^{(p+1)/2} 
      |V_{k,i}|^{1/2}} \exp \left\{ -\frac{1}{2} ({\bf z}_i - \zeta_k)^T V_{k,i}^{-1} 
      ({\bf z}_i - \zeta_k) \right\} \label{eq-mobslik} \\
    \zeta_k & = & (\alpha + \beta^T \mu_k, \mu_k) \label{eq-mzeta} \\
    V_{k,i} & = & \left( \begin{array}{cc}
      \beta^T T_k \beta + \sigma^2 + \sigma^2_{y,i} & 
      \beta^T T_k + \sigma^T_{xy,i} \\
      T_k \beta + \sigma_{xy,i} & T_k + \Sigma_{x,i} \end{array} \right).
    \label{eq-msigmaz}
  \end{eqnarray}
  Here, $\zeta_k$ is the $(p+1)$-element mean vector of ${\bf z}_i$
  for Gaussian $k$, $V_{k,i}$ is the $(p+1) \times (p+1)$ covariance
  matrix of ${\bf z}_i$ for Gaussian $k$, $\sigma^2_{y,i}$ is the
  variance in the measurement error on $y_i$, $\sigma_{xy,i}$ is the
  $p$-element vector of covariances between the measurement errors on
  $y_i$ and ${\bf x}_i$, and $\Sigma_{x,i}$ is the $p \times p$
  covariance matrix of the measurement errors on ${\bf x}_i$.

  Similar to the case for one independent variable, the measured data
  likelihood can be decomposed as $p(x,y|\theta,\psi) =
  p(y|x,\theta,\psi) p(x|\psi)$, where $p({\bf x}_i|\psi) =
  \sum_{k=1}^K \pi_k N_p({\bf x}_i|\mu_k, T_k + \Sigma_{x,i})$ and
  \begin{eqnarray}
    p(y_i|{\bf x}_i, \theta, \psi) & = & \sum_{k=1}^K 
    \frac{\gamma_k}{\sqrt{2\pi Var(y_i|{\bf x}_i,k)}}
    \exp \left\{ -\frac{1}{2} \frac{[y_i - E(y_i|{\bf x}_i,k)]^2}{Var(y_i|{\bf x}_i,k)} 
      \right\} \label{eq-mcdensyx} \\
    \gamma_k & = &  \frac{\pi_k N({\bf x}_i|\mu_k, T_k + \Sigma_{x,i})}
	  {\sum_{j=1}^K \pi_j N({\bf x}_i|\mu_j, T_j + \Sigma_{x,i})}
	  \label{eq-mcdenskx} \\
    E(y_i|{\bf x}_i,k) & = & \alpha + \beta^T \mu_k + (\beta^T T_k + \sigma^T_{xy,i})
    (T_k + \Sigma_{x,i})^{-1} ({\bf x}_i - \mu_k) \label{eq-mxihati} \\
    Var(y_i|{\bf x}_i, k) & = & \beta^T T_k \beta + \sigma^2 + \sigma^2_{y,i} - 
    (\beta^T T_k + \sigma^T_{xy,i}) (T_k + \Sigma_{x,i})^{-1} (T_k \beta + \sigma_{xy,i})
    \label{eq-mcvary}.
  \end{eqnarray}

  \section{DATA COLLECTION ISSUES: SELECTION EFFECTS AND NON-DETECTIONS}

  \label{s-dcissues}

  There are several issues common in the collection of astronomical
  data that violate the simple assumptions made in \S~\ref{s-statmod}.
  Astronomical data collection consists almost entirely of passive
  observations, and thus selection effects are a common
  concern. Instrumental detection limits often result in the placement
  of upper or lower limits on quantities, and astronomical surveys are
  frequently flux-limited. In this section I modify the likelihood
  functions described in \S~\ref{s-statmod} to include the effects of
  data collection.

  General methods for dealing with missing data are described in
  \citet{lit02} and \citet{gelman04}, and I apply the methodology
  described in these references to the measurement error model
  developed here. Although in this work I focus on linear regression,
  many of these results can be applied to more general statistical
  models, such as estimating luminosity functions.

  \subsection{Selection Effects}

  \label{s-seffects}

  Suppose that one collects a sample of $n$ sources out of a possible
  $N$ sources. One is interested in understanding how the observable
  properties of these sources are related, but is concerned about the
  effects of the selection procedure on the data analysis. For
  example, one may perform a survey that probes some area of the
  sky. There are $N$ sources located within this solid angle, where
  $N$ is unknown. Because of the survey's selection method, the sample
  only includes $n$ sources. In this case the astronomer is interested
  in how measurement error and the survey's selection method affect
  statistical inference.

  I investigate selection effects within the framework of our
  statistical model by introducing an indicator variable, $I$, which
  denotes whether a source is included in the sample. If the $i^{\rm
  th}$ source is included in the sample, then $I_i = 1$, otherwise
  $I_i = 0$. In addition, I assume that the selection function only
  depends on the measured values, $x$ and $y$. Under this assumption,
  the selection function of the sample is the probability of including
  a source with a given $x$ and $y$, $p(I|x,y)$. This is commonly the
  case in astronomy, where sources are collected based on their
  measured properties. For example, one may select sources for a
  sample based on their measured properties as reported in the
  literature. In addition, if one performs a flux-limited survey then
  a source will only be considered detected if its measured flux falls
  above some set flux limit.  If a sample is from a survey with a
  simple flux limit, then $p(I_i=1|y_i) = 1$ if the measured source
  flux $y_i$ is above the flux limit, and $p(I_i=1|y_i) = 0$ if the
  measured source flux is below the flux limit. Since the selection
  function depends on the measured flux value, and not the true flux
  value, sources with true flux values above the flux limit can be
  missed by the survey, and sources with true flux below the limit can
  be detected by the survey. This effect is well-known in astronomy
  and is commonly referred to as Malmquist bias
  \citep[e.g.,][]{landy92}.

  Including the variable $I$, the complete data likelihood can be
  written as
  \begin{equation}
    p(x,y,\xi,\eta,I|\theta,\psi) = p(I|x,y) p(x,y|\xi,\eta)
    p(\eta|\xi,\theta) p(\xi|\psi).
    \label{eq-complik}
  \end{equation}
  Equation (\ref{eq-complik}) is valid for any number of independent
  variables, and thus $x_i$ and $\xi_i$ may be either scalar or
  vector. Integrating Equation (\ref{eq-complik}) over the missing
  data, the observed data likelihood is
  \begin{eqnarray}
    \lefteqn{p(x_{obs}, y_{obs}|\theta, \psi, N) \propto \binomial \prod_{i \in {\cal A}_{obs}}
      p(x_i, y_i|\theta, \psi)} \nonumber \\
    & & \times \prod_{j \in {\cal A}_{mis}} 
    \int p(I_j = 0|x_j,y_j) p(x_j,y_j|\xi_j,\eta_j) p(\eta_j|\xi_j,\theta) p(\xi_j|\psi)
    \ dx_j\ dy_j\ d\xi_j\ d\eta_j.
    \label{eq-mislik}
  \end{eqnarray}
  Here, $\binomial$ is the binomial coefficient, ${\cal A}_{obs}$
  denotes the set of $n$ included sources, $x_{obs}$ and $y_{obs}$
  denote the values of $x$ and $y$ for the included sources, and
  ${\cal A}_{mis}$ denotes the set of $N - n$ missing sources. In
  addition, I have omitted terms that do not depend on $\theta,
  \psi$, or $N$. Note that $N$ is unknown and is thus also a parameter
  of the statistical model. The binomial coefficient is necessary
  because it gives the number of possible ways to select a sample of
  $n$ sources from a set of $N$ sources.

  It is apparent from Equation (\ref{eq-mislik}) that statistical
  inference on the regression parameters is unaffected if the
  selection function is independent of $y$ and $x$.
  \citep[e.g.,][]{lit02,gelman04}. In this case the selection function
  may be ignored.

  \subsubsection{Selection Based on Measured Independent Variables}

  \label{s-seffects_x}

  It is commonly the case that a sample is selected based only on the
  measured independent variables. For example, suppose one performs a
  survey in which all sources with measured optical flux greater than
  some threshold are included. Then, these optically selected sources
  are used to fit a regression in order to understand how the X-ray
  luminosity of these objects depends on their optical luminosity and
  redshift. In this case, the probability of including a source only
  depends on the measured values of the optical luminosity and
  redshift, and is thus independent of the X-ray luminosity.

  When the sample selection function is independent of $y$, given $x$,
  then $p(I|x,y) = p(I|x)$. Because we are primarily interested in the
  regression parameters, $\theta$, I model the distributions of $\xi$
  for the included and missing sources seperately, with the parameters
  for the distribution of included sources denoted as $\psi_{obs}$. In
  addition, I assume that the measurement errors between $y$ and $x$
  are statistically independent. Then the $N - n$ integrals over $y$
  and $\eta$ for the missing sources in Equation (\ref{eq-mislik}) are
  equal to unity, and we can write the observed data likelihood as
  \begin{equation}
    p(x_{obs}, y_{obs}|\theta,\psi_{obs}) \propto \prod_{i=1}^{n} \int
    \int p(x_i|\xi_i) p(y_i|\eta_i) p(\eta_i|\xi_i,\theta)
    p(\xi_i|I_i=1,\psi_{obs})\ d\xi_i\ d\eta_i,
    \label{eq-indlik}
  \end{equation}
  where $p(\xi_i|I_i=1,\psi_{obs})$ is the distribution of those $\xi$
  included in one's sample. Here I have omitted terms depending on
  $N$ because one is primarily interested in inference on the
  regression parameters, $\theta$.  Equation (\ref{eq-indlik}) is
  identical to Equation (\ref{eq-obslik2}), with the exception that
  $p(\xi|\psi)$ now only models the distribution of those $\xi$ that
  have been included in one's sample, and I have now assumed that the
  measurement errors on $y$ and $x$ are independent. In particular,
  for the Gaussian mixture models described in \S~\ref{s-normreg} and
  \S~\ref{s-multreg}, the observed data likelihood is given by
  Equations (\ref{eq-bobslik}) and (\ref{eq-mobslik}), where $\pi$,
  $\mu$, and $\tau^2$ (or $T$) should be understood as referring to
  the parameters for the distribution of the observed $\xi$. As is
  evident from the similarity between Equations (\ref{eq-indlik}) and
  (\ref{eq-obslik2}), \emph{if the sample is selected based on the
  measured independent variables, and if the measurement errors on the
  dependent and independent variables are statistically independent,
  then inference on the regression parameters, $\theta$, is unaffected
  by selection effects}.

  \subsubsection{Selection Based on Measured Dependent and Independent Variables}

  \label{s-seffects_xy}

  If the method in which a sample is selected depends on the measured
  dependent variable, $y$, or if the measurement error in $x$ and $y$
  are correlated, the observed data likelihood becomes more
  complicated. As an example, one might encounter this situation if
  one uses an X-ray selected sample to investigate the dependence of
  X-ray luminosity on optical luminosity and redshift. In this case,
  the selection function of the sample depends on both the X-ray
  luminosity and redshift, and is thus no longer independent of the
  dependent variable. Such data sets are said to be `truncated'.

  If the selection function depends on $y$, or if the measurement
  errors on $y$ and $x$ are not independent, one cannot simply ignore
  the terms depending on $N$, since the $N-n$ integrals in Equation
  (\ref{eq-mislik}) depend on $\theta$.  However, we can eliminate the
  dependence of Equation (\ref{eq-mislik}) on the unknown $N$ by
  applying a Bayesian approach. The posterior distribution of $\theta,
  \psi,$ and $N$ is related to the observed data likelihood function
  as $p(\theta,\psi,N|x_{obs},y_{obs}) \propto p(\theta,\psi,N)
  p(x_{obs},y_{obs}|\theta,\psi,N)$, where $p(\theta,\psi,N)$ is the
  prior distribution of $(\theta,\psi,N)$. If we assume a uniform
  prior on $\theta, \psi,$ and $\log N$, then one can show
  \citep[e.g.,][]{gelman04} that the posterior distribution of
  $\theta$ and $\psi$ is
  \begin{equation}
    p(\theta,\psi|x_{obs},y_{obs}) \propto \left[ p(I=1|\theta,\psi) \right]^{-n} 
    \prod_{i=1}^n p(x_i,y_i|\theta, \psi).
    \label{eq-trunclik}
  \end{equation}
  Here, $p(x_i,y_i|\theta,\psi)$ is given by Equation
  (\ref{eq-obslik2}), and $p(I=1|\theta,\psi)$ is the probability of
  including a source in one's sample, given the model parameters,
  $\theta$ and $\psi$:
  \begin{equation}
    p(I=1|\theta,\psi) = \int \int p(I = 1|x,y) p(x,y|\theta,\psi) \ dx\ dy.
    \label{eq-detprob}
  \end{equation}
  I have left off the subscripts for the data points in Equation
  (\ref{eq-detprob}) because the integrals are the same for each
  $(x_j,y_j,\xi_j,\eta_j)$. If one assumes the Gaussian mixture model
  of Sections \ref{s-normreg} and \ref{s-multreg}, then
  $p(x_i,y_i|\theta,\psi)$ is given by Equations (\ref{eq-bobslik}) or
  (\ref{eq-mobslik}). The posterior mode can then be used as an
  estimate of $\theta$ and $\psi$, which is found by maximizing
  Equation (\ref{eq-trunclik}).

  \subsection{Non-detections}

  \label{s-nondet}

  In addition to issues related to the sample selection method, it is
  common in astronomical data to have non-detections. Such
  non-detections are referred to as `censored' data, and the standard
  procedure is to place an upper and/or lower limit on the censored
  data point. Methods of data analysis for censored data have been
  reviewed and proposed in the astronomical literature,
  \citep[e.g.,][]{feig85,schmitt85,marsh92,akritas96}, and
  \citet{isobe86} describe censored regression when the variables are
  measured without error. See \citet{feig92} for a review of censored
  data in astronomy.

  To facilitate the inclusion of censored data, I introduce an
  additional indicator variable, $D$, indicating whether a data point
  is censored or not on the dependent variable. If $y_i$ is detected,
  then $D_i = 1$, else if $y_i$ is censored then $D_i = 0$. It is
  commonly the case that a source is considered `detected' if its
  measured flux falls above some multiple of the background noise
  level, say $3\sigma$. Then, in this case, the probability of
  detecting the source given the measured source flux $y_i$ is
  $p(D_i=1|y_i) = 1$ if $y_i > 3\sigma$, and $p(D_i=0|y_i) = 1$ if
  $y_i < 3\sigma$. Since source detection depends on the measured
  flux, some sources with intrinsic flux $\eta$ above the flux limit
  will have a measured flux $y$ that falls below the flux
  limit. Similarly, some sources with intrinsic flux below the flux
  limit will have a measured flux above the flux limit.

  I assume that a sample is selected based on the independent
  variables, i.e., $p(I|x,y) = p(I|x)$. It is difficult to imagine
  obtaining a censored sample if the sample is selected based on its
  dependent variable, as some of the values of $y$ are censored and
  thus unknown. Therefore, I only investigate the effects of
  censoring on $y$ when the probability that a source is included in
  the sample is independent of $y$, given $x$. In addition, I do not
  address the issue of censoring on the independent variable. Although
  such methods can be developed, it is probably simpler to just omit
  such data as inference on the regression parameters is unaffected
  when a sample is selected based only on the independent variables
  (cf., \S~\ref{s-seffects_x}).

  The observed data likelihood for an $x$-selected sample is given by
  Equation (\ref{eq-indlik}). We can modify this likelihood to account
  for censored $y$ by including the indicator variable $D$ and again
  integrating over the missing data:
  \begin{equation}
    p(x_{obs},y_{obs},D|\theta,\psi_{obs}) \propto \prod_{i \in {\cal A}_{det}} 
    p(x_i, y_i|\theta, \psi_{obs}) \prod_{j \in {\cal A}_{cens}} p(x_j|\psi_{obs}) 
    \int p(D_j=0|y_j,x_j) p(y_j|x_j, \theta,\psi_{obs})\ dy_j.
    \label{eq-censlik}
  \end{equation}
  Here, the first product is over the set of data points with
  detections, ${\cal A}_{det}$, and the second product is over the set
  of data points with non-detections, ${\cal A}_{cens}$. The
  conditional distribution $p(y_j|x_j,\theta,\psi_{obs})$ and the
  marginal distribution $p(x_j|\psi_{obs})$ for the Gaussian mixture
  model are both given in \S~\ref{s-normreg} and
  \S~\ref{s-multreg}. If the data points are measured without error
  and one assumes the normal regression model, $p(\eta|\xi,\theta) =
  N(\eta|\alpha + \beta \xi,\sigma^2)$, then Equation \ref{eq-censlik}
  becomes the censored data likelihood function described in
  \citet{isobe86}. A MLE for censored regression with measurement
  errors is then obtained by maximizing Equation (\ref{eq-censlik}).

  \section{COMPUTATIONAL METHODS}

  \label{s-compmeth}

  In this section I describe a Bayesian method for computing
  estimates of the regression parameters, $\theta$, and their
  uncertainties. The Bayesian approach calculates the posterior
  probability distribution of the model parameters, given the observed
  data, and therefore is accurate for both small and large sample
  sizes. The posterior distribution follows from Baye's formula as
  $p(\theta,\psi|x,y) \propto p(\theta,\psi) p(x,y|\theta,\psi)$,
  where $p(\theta,\psi)$ is the prior distribution of the parameters.
  I describe some Markov Chain methods for drawing random variables
  from the posterior, which can then be used to estimate quantities
  such as standard errors and confidence intervals on $\theta$ and
  $\psi$. \citet{gelman04} is a good reference on Bayesian methods,
  and \citet{loredo92} gives a review of Bayesian methods intended for
  astronomers. Further details of Markov Chain simulation, including
  methods for making the simulations more efficient, can be found in
  \citet{gelman04}.

  \subsection{The Prior Density}

  \label{s-prior}

  In order to ensure a proper posterior for the Gaussian mixture
  model, it is necessary to invoke a proper prior density on the
  mixture parameters \citep{roeder97}. I adopt a uniform prior on the
  regression parameters $(\alpha, \beta, \sigma^2)$, and take
  $\pi_1,\ldots,\pi_K \sim {\rm Dirichlet}(1,\ldots,1)$. The Dirichlet
  density is a multivariate extension of the Beta density, and the
  ${\rm Dirichlet}(1,\ldots,1)$ prior adopted in this work is
  equivalent to a uniform prior on $\pi$, under the constraint
  $\sum_{k=1}^K \pi_k = 1$.

  The prior on $\mu$ and $\tau^2$ (or $T$) adopted in this work is
  very similar to that advocated by \citet{roeder97} and
  \citet{carroll99}. I adopt a normal prior on the individual $\mu_k$
  with mean $\mu_0$ and variance $u^2$ (or covariance matrix
  $U$). This reflects our prior belief that the distribution of $\xi$
  is more likely to be fairly unimodal, and thus that we expect it to
  be more likely that the individual Gaussians will be close together
  than far apart. If there is only one covariate, then I adopt a
  scaled inverse-$\chi^2$ prior on the individual $\tau_k^2$ with
  scale parameter $w^2$ and one degree of freedom, otherwise if there
  are $p > 1$ covariates I adopt an inverse-Wishart prior on the
  individual $T_k$ with scale matrix $W$ and $p$ degrees of
  freedom. This reflects our prior expectation that the variances for
  the individual Gaussian components should be similar, but the low
  number of degrees of freedom accomodates a large range of
  scales. Both the Gaussian means and variances are assumed to be
  independent in their prior distribution, and the `hyper-parameters'
  $\mu_0, u^2$ (or $U$), and $w^2$ (or $W$) are left unspecified. By
  leaving the parameters for the prior distribution unspecified, they
  becomes additional parameters in the statistical model, and
  therefore are able to adapt to the data.
  
  Since the hyper-parameters are left as free parameters they also
  require a prior density. I assume a uniform prior on $\mu_0$ and
  $w^2$ (or $W$). If there is one covariate, then I assume a scaled
  inverse-$\chi^2$ prior for $u^2$ with scale parameter $w^2$ and one
  degree of freedom, otherwise if there are multiple covariate we
  assume a inverse-Wishart prior for $U$ with scale matrix $W$ and $p$
  degrees of freedom. The prior on $u^2$ (or $U$) reflects the prior
  expectation that the dispersion of the Gaussian components about
  their mean $\mu_0$ should be on the order of the typical dispersion
  of each individual Gaussian. The prior density for one covariate is
  then $p(\theta,\psi,\mu_0,u^2,w^2) \propto p(\pi) p(\mu|\mu_0,u^2)
  p(\tau^2|w^2) p(u^2|w^2)$ and is summarized hierarchically as
  \begin{eqnarray}
    \alpha,\beta & \sim & {\rm Uniform}(-\infty,\infty) \label{eq-coefprior} \\
    \sigma^2 & \sim & {\rm Uniform}(0,\infty) \label{eq-ssqrprior} \\
    \pi & \sim & {\rm Dirichlet}(1,\ldots,1) \label{eq-piprior} \\
    \mu_1,\ldots,\mu_K|\mu_0,u^2 & \sim & N(\mu_0,u^2) \label{eq-muprior} \\
    \tau^2_1,\ldots,\tau^2_K,u^2|w^2 & \sim & \mbox{\rm Inv-}\chi^2(1,w^2) \label{eq-tsqrprior} \\
    \mu_0 & \sim & {\rm Uniform}(-\infty,\infty) \label{eq-mu0prior} \\
    w^2 & \sim & {\rm Uniform}(0,\infty) \label{eq-wsqrprior}.
  \end{eqnarray}
  The prior density for multiple covariates is just the multivariate
  extension of Equations (\ref{eq-coefprior})--(\ref{eq-wsqrprior}).

  \subsection{Markov Chains for Sampling from the Posterior Distribution}

  \label{s-markov}

  The posterior distribution summarizes our knowledge about the
  parameters in the statistical model, given the observed data and the
  priors. Direct computation of the posterior distribution is too
  computationally intensive for the model described in this work.
  However, we can obtain any number of random draws from the posterior
  using Markov chain monte carlo (MCMC) methods. In MCMC methods, we
  simulate a Markov chain that performs a random walk through the
  parameter space, saving the locations of the walk at each
  iteration. Eventually, the Markov chain converges to the posterior
  distribution, and the saved parameter values can be treated as a
  random draw from the posterior. The random draws can then be used to
  estimate posterior medians, standard errors, of plot histogram
  estimates of the posterior.

  \subsubsection{Gibbs Sampler for the Gaussian Model}

  \label{s-gibbs}

  The easiest method for sampling from the posterior is to construct a
  Gibbs sampler. The basic idea behind the Gibbs sampler is to
  construct a Markov Chain, where new values of the model parameters
  and missing data are simulated at each iteration, conditional on the
  values of the observed data and the current values of the model
  parameters and the missing data. Within the context of the
  measurement error model considered in this work, the Gibbs Sampler
  undergoes four different stages.

  The first stage of the Gibbs sampler simulates values of the missing
  data, given the measured data and current parameter values, a
  process known as data augmentation. In this work the missing data
  are $\eta, \xi,$ and any non-detections. In addition, I introduce
  an additional latent variable, ${\bf G}_i$, which gives the class
  membership for the $i^{\rm th}$ data point. The vector ${\bf G}_i$
  has $K$ elements, where $G_{ik} = 1$ if the $i^{\rm th}$ data point
  comes from the $k^{\rm th}$ Gaussian, and $G_{ij} = 0$ if $j \neq
  k$. I will use $G$ to refer to the set of $n$ vectors ${\bf
  G}_i$. Noting that $\pi_k$ gives the probability of drawing a data
  point from the $k^{\rm th}$ Gaussian, the mixture model for $\xi$
  may then be expressed hierarchically as
  \begin{eqnarray}
    {\bf G}_i|\pi & \sim & {\rm Multinom}(1,\pi_1,\ldots,\pi_K) \label{eq-hierg} \\
    \xi_i|G_{ik} = 1, \mu_k, \tau_k^2 & \sim & N(\mu_k,\tau^2_k). \label{eq-hierxi2}
  \end{eqnarray}
  Here, ${\rm Multinom}(m,p_1,\ldots,p_K)$ is a multinomial
  distribution with $m$ trials, where $p_k$ is the probability of
  success for the $k^{\rm th}$ class on any particular trial. The
  vector ${\bf G}_i$ is also considered to be missing data, and is
  introduced to simplify construction of the Gibbs sampler.

  The new values of the missing data simulated in the data
  augmentation step are then used to simulate new values of the
  regression and Gaussian mixture parameters. The second stage of the
  Gibbs sampler simulates values of the regression parameters,
  $\theta$, given the current values of $xi$ and $\eta$. The third
  stage simulates values of the mixture parameters, $\psi$, given the
  current values of $\xi$ and $\eta$. The fourth stage uses the new
  values of $\theta$ and $\psi$ to update the parameters of the prior
  density. The values of the parameters are saved, and the process is
  repeated, creating a Markov Chain. After a large number of
  iterations, the Markov Chain converges, and the saved values of
  $\theta$ and $\psi$ from the latter part of the algorithm may then
  be treated as a random draw from the posterior distribution,
  $p(\theta,\psi|x,y)$.

  Methods for simulating random variables from the distributions used
  for this Gibbs sampler are described in various works
  \citep[e.g.,][]{rip87,numrec,gelman04}.

  A Gibbs sampler for the Gaussian mixture model is
  \begin{enumerate}
  \item
    \label{i-gstart}
    Start with initial guesses for $\eta, G, \theta, \psi,$ and the
    prior parameters.
  \item
    \label{i-ycens}
    If there are any non-detections, then draw $y_i$ for the censored
    data points from $p(y_i|\eta_i,D_i=0) \propto p(D_i=0|y_i)
    p(y_i|\eta_i)$. This may be done by first drawing $y_i$ from
    $p(y_i|\eta_i)$:
    \begin{equation}
      y_i|\eta_i \sim N(\eta_i,\sigma^2_{y,i}) \label{eq-gycens}.
    \end{equation}
    One then draws a random variable $u_i$, uniformly-distributed on
    $[0,1]$. If $u_i < p(D_i=0|y_i)$ then the value of $y_i$ is kept,
    otherwise one draws a new value of $y_i$ and $u_i$ until $u_i <
    p(D_i=0|y_i)$.
  \item
    \label{i-xi}
    Draw values of $\xi$ from $p(\xi|x,y,\eta,G,\theta,\psi)$. The
    distribution $p(\xi|x,y,\eta,G,\theta,\psi)$ can be derived from
    Equations (\ref{eq-hierxi})--(\ref{eq-hierxy}) or
    (\ref{eq-mhierxi})--(\ref{eq-mhierxy}) and the properties of the
    multivariate normal distribution:
    \begin{enumerate}
    \item
      \label{i-xi1}
      If there is only one independent variable then $\xi_i$ is updated as:
      \begin{eqnarray}
	\xi_i|x_i,y_i,\eta_i,{\bf G}_i,\theta,\psi & \sim & N(\hat{\xi}_i, \sigma^2_{\hat{\xi},i}) 
           \label{eq-xicond} \\
	\hat{\xi}_i & = & \sum_{k=1}^{K} G_{ik} \hat{\xi}_{ik} \label{eq-gxihat} \\
	\hat{\xi}_{ik} & = & \sigma^2_{\hat{\xi},i} \left[ \frac{\hat{\xi}_{xy,i}}
	    {\sigma^2_{x,i}(1 - \rho^2_{xy,i})} + \frac{\beta(\eta_i - \alpha)}{\sigma^2} + 
	    \frac{\mu_k}{\tau_k^2} \right] 
	    \label{eq-gxihatk} \\
	\hat{\xi}_{xy,i} & = & x_i + \frac{\sigma_{xy,i}}{\sigma^2_{y,i}} (\eta_i - y_i)
	   \label{eq-gxixy} \\
	\sigma^2_{\hat{\xi},i} & = & \sum_{k=1}^{K} G_{ik} \sigma^2_{\hat{\xi},ik} 
	   \label{eq-gxihvark} \\
	\sigma^2_{\hat{\xi},ik} & = & \left[ \frac{1}{\sigma^2_{x,i}(1 - \rho^2_{xy,i})} + 
           \frac{\beta^2}{\sigma^2} + \frac{1}{\tau_k^2} \right]^{-1}. \label{eq-gxihvar}
    \end{eqnarray}
      Here, $\rho_{xy,i} = \sigma_{xy,i} / (\sigma_{x,i}
      \sigma_{y,i})$ is the correlation between the measurement errors
      on $x_i$ and $y_i$. Note that $\xi_i$ is updated using only
      information from the $k^{\rm th}$ Gaussian, since $G_{ij} = 1$
      only for $j=k$ and $G_{ij} = 0$ otherwise. 
    \item
      \label{i-xi2}
      If there are multiple independent variables, I have found it
      easier and computationally faster to update the values of
      $\xi_i$ using a scalar Gibbs sampler. In this case, the $p$
      elements of $\xi_i$ are updated individually. I denote
      $\xi_{ij}$ to be the value of the $j^{\rm th}$ independent
      variable for the $i^{\rm th}$ data point, and $x_{ij}$ to be the
      measured value of $\xi_{ij}$. In addition, I denote
      $\xi_{i,-j}$ to be the $(p-1)$-element vector obtained by
      removing $\xi_{ij}$ from $\xi_i$, i.e., $\xi_{i,-j} =
      (\xi_{i1}, \ldots, \xi_{i(j-1)}, \xi_{i(j+1)}, \ldots,
      \xi_{ip})$. Similarly, $\beta_{-j}$ denotes the $(p-1)$-element
      vector of regression coefficients obtained after removing
      $\beta_j$ from $\beta$. Then, $\xi_{ij}$ is updated as
      \begin{eqnarray}
	\xi_{ij} | {\bf x}_i, y_i, {\bf G}_i, \xi_{i,-j}, \eta_i, \theta, \psi & \sim & 
            N(\hat{\xi}_{ij},\sigma^2_{\hat{\xi},ij}) \label{eq-mxicond} \\
	\hat{\xi}_{ij} & = & \sum_{k=1}^K G_{ik} \hat{\xi}_{ijk} \label{eq-gmxihat} \\
	\hat{\xi}_{ijk} & = & \frac{ (\Sigma^{-1}_i {\bf z}_i^*)_{j+1} + 
	  (T_k^{-1} \mu^*_{ik})_j + \beta_j (\eta_i - \alpha - \beta^T_{-j} \xi_{i,-j}) / 
	  \sigma^2}{ (\Sigma^{-1}_i)_{(j+1)(j+1)} + (T^{-1}_k)_{jj} + \beta^2_j / \sigma^2}
          \label{eq-gmxihatk} \\
	({\bf z}^*_i)_l & = & \left\{ \begin{array}{l}
	  y_i - \eta_i\ {\rm if}\ l = 1 \\
	  x_{il}\ {\rm if}\ l = j+1 \\
	  x_{il} - \xi_{il}\ {\rm if}\ l \neq j+1
	\end{array} \right. \label{eq-xstar} \\
	(\mu^*_{ik})_l & = & \left\{ \begin{array}{l}
	  (\mu_k)_l\ {\rm if}\ l = j \\
	  (\mu_k)_l - \xi_{il}\ {\rm if}\ l \neq j
	\end{array} \right. \label{eq-mustar} \\
	\sigma^2_{\hat{\xi},ij} & = & \sum_{k=1}^K G_{ik} \sigma^2_{\hat{\xi},ijk} 
	  \label{eq-gmxihvar} \\
	\sigma^2_{\hat{\xi},ijk} & = & \left[ (\Sigma^{-1}_i)_{(j+1)(j+1)} + 
	  (T^{-1}_k)_{jj} + \frac{\beta^2_j}{\sigma^2} \right]^{-1}. 
        \label{eq-gmxihvark}
      \end{eqnarray}
      Here, ${\bf z}^*_i$ is a $(p+1)$-element vector obtained by
      subtracting $(\eta_i,\xi_i)$ from ${\bf z}_i = (y_i,{\bf x}_i)$,
      with the exception of the $j^{\rm th}$ element of $\xi_i$;
      instead, the $(j+1)^{\rm th}$ element of ${\bf z}_i^*$ is just
      $x_{ij}$. The $p$-element vector $\mu^*_{ik}$ is obtained in an
      equivalent manner. The $(p+1) \times (p+1)$ matrix $\Sigma_i$ is
      the covariance matrix of the measurement errors on ${\bf
      z}_i$. The term $(\Sigma^{-1}_i {\bf z}^*_i)_{(j+1)}$ denotes the
      $(j+1)^{\rm th}$ element of the vector $\Sigma^{-1}_i {\bf
      z}^*_i$, and likewise for $(T_k^{-1} \mu^*_{ik})_j$. The terms
      $(\Sigma^{-1}_i)_{(j+1)(j+1)}$ and $(T^{-1}_k)_{jj}$ denote the
      $(j+1)^{\rm th}$ and $j^{\rm th}$ elements of the diagonals of
      $\Sigma^{-1}_i$ and $T_k^{-1}$, respectively. This step is
      repeated until all $p$ independent variables have been updated
      for each data point.
    \end{enumerate}
    If any of the $\xi_i$ are measured without error, then one simply
    sets $\xi_i = x_i$ for those data points.
  \item
    \label{i-eta}
    Draw values of $\eta$ from $p(\eta|x,y,\xi,\theta)$. Similar to
    $\xi$, the distribution $p(\eta|x,y,\xi,\theta)$ can be derived from
    Equations (\ref{eq-hierxi})--(\ref{eq-hierxy}) or
    (\ref{eq-mhierxi})--(\ref{eq-mhierxy}) and the properties of the
    multivariate normal distribution.
    \begin{enumerate}
    \item
      \label{i-eta1}
      If there is only one covariate then $\eta$ is updated as
      \begin{eqnarray}
	\eta_i|x_i,y_i,\xi_i,\theta & \sim & N(\hat{\eta}_i, \sigma^2_{\hat{\eta},i})
        \label{eq-metacond} \\
	\hat{\eta}_i & = & \sigma^2_{\hat{\eta},i} 
	\left[ \frac{y_i + \sigma_{xy,i}(\xi_i - x_i) / 
	  \sigma^2_{x,i}}{\sigma^2_{y,i}(1 - \rho^2_{xy,i})} + 
        \frac{\alpha + \beta \xi_i}{\sigma^2} \right] \label{eq-gmetahat} \\
	\sigma^2_{\hat{\eta},i} & = &  \left[ \frac{1}{\sigma^2_{y,i}(1 - \rho^2_{xy,i})} + 
        \frac{1}{\sigma^2} \right]^{-1}. \label{eq-gmetahvar}
      \end{eqnarray}
    \item
      \label{i-eta2}
      If there are multiple covariates then $\eta$ is updated as
      \begin{eqnarray}
	\eta_i|{\bf x}_i,y_i,\xi_i,\theta & \sim & N(\hat{\eta}_i, \sigma^2_{\hat{\eta},i})
	\label{eq-metacond2} \\
	\hat{\eta}_i & = & \frac{(\Sigma^{-1}_i {\bf z}^*_i)_1 + 
	  (\alpha + \beta^T \xi_i) / \sigma^2}{(\Sigma^{-1}_i)_{11} + 1 / \sigma^2}
	\label{eq-gmetahat2} \\
	\sigma^2_{\hat{\eta},i} & = & \left[ (\Sigma^{-1}_i)_{11} + 
	\frac{1}{\sigma^2} \right]^{-1} \label{eq-gmetahvar2} \\
	{\bf z}^*_i & = & (y_i, {\bf x}_i - \xi_i). \label{eq-zstar}
      \end{eqnarray}
      Here, $(\Sigma_i^{-1} {\bf z}_i^*)_1$ is the first element of the
      vector $\Sigma_i^{-1} {\bf z}_i^*$, ${\bf z}^*_i$ is a
      $(p+1)$-element vector whose first element is $y_i$ and remaining
      elements are ${\bf x}_i - \xi_i$, and $(\Sigma^{-1}_i)_{11}$ is
      the first diagonal element of $\Sigma^{-1}_i$. 
    \end{enumerate}
    If any of the
    $\eta$ are measured without error, then one sets $\eta = y$ for
    those data points.
  \item
    \label{i-glabel}
    Draw new values of the Gaussian labels, $G$. The conditional
    distribution of ${\bf G}_i$ is Multinomial with number of trials
    $m = 1$ and group probabilities $q_k = p(G_{ik} = 1|\xi_i,\psi)$:
    \begin{eqnarray}
      {\bf G}_i|\xi_i,\psi & \sim & {\rm Multinom}(1,q_1,\ldots,q_K) \label{eq-glabel} \\
      q_k & = & \frac{\pi_k N_p (\xi_i|\mu_k,T_k)}{\sum_{j=1}^K \pi_j 
	N_p(\xi_i|\mu_j,T_j)}. \label{eq-gclassprob}
    \end{eqnarray}
    Note that if there is only one covariate then $p = 1$ and $T_k = \tau^2_k$.
  \item
    \label{i-coef}
    Draw $(\alpha, \beta)$ from $p(\alpha,\beta|\xi, \eta,
    \sigma^2)$. Given $\xi, \eta,$ and $\sigma^2$, the distribution of
    $\alpha$ and $\beta$ is obtained by ordinary regression:
    \begin{eqnarray}
      \alpha, \beta|\xi,\eta,\sigma^2 & \sim & N_{p+1} (\hat{\bf c}, \Sigma_{\hat{\bf c}}) 
        \label{eq-coefcond} \\
      \hat{\bf c} & = & (X^T X)^{-1} X^T \eta \label{eq-chat} \\
      \Sigma_{\hat{\bf c}} & = & (X^T X)^{-1} \sigma^2. \label{eq-sigchat}
    \end{eqnarray}
    Here, $X$ is a $n \times (p+1)$ matrix, where the first
    column is a column of ones, the second column contains the $n$
    values of $\xi_i$ for the first independent variable, the third
    column contains the $n$ values of $\xi_i$ for the second
    independent variable, etc.
  \item
    \label{i-sigsqr}
    Draw a new value of $\sigma^2$ from
    $p(\sigma^2|\xi,\eta,\alpha,\beta)$. The distribution
    $p(\sigma^2|\xi,\eta,\alpha,\beta)$ is derived by noting that
    given $\alpha,\beta$ and $\xi_i$, $\eta_i$ is normally distributed
    with mean $\alpha + \beta^T \xi_i$ and variance
    $\sigma^2$. Re-expressing this distribution in terms of $\sigma^2$
    instead of $\eta$, and taking the product of the distributions for
    each data point, it follows that $\sigma^2$ has a scaled
    inverse-$\chi^2$ distribution:
    \begin{eqnarray}
      \sigma^2|\xi,\eta,\alpha,\beta & \sim & \mbox{\rm Inv-}\chi^2(\nu,s^2) \label{eq-sigcond1} \\
      \nu & = & n - 2 \label{eq-snu1} \\
      s^2 & = & \frac{1}{n-2} \sum_{i=1}^n (\eta_i - \alpha - \beta^T \xi_i)^2. \label{eq-ssqr1}
    \end{eqnarray}
  \item
    \label{i-pi}
    Draw new values of the group proportions, $\pi$. Given $G$, $\pi$
    follows a Dirichlet distribution:
    \begin{eqnarray}
      \pi|G & \sim & {\rm Dirichlet}(n_1+1,\ldots,n_K+1) \label{eq-gpi} \\
      n_k & = & \sum_{i=1}^n G_{ik}. \label{eq-gnk}
    \end{eqnarray}
    Note that $n_k$ is the number of data points that belong to the
    $k^{\rm th}$ Gaussian.
  \item
    \label{i-mu}
    Draw a new value of $\mu_k$ from $p(\mu_k|\xi,G,T_k,\mu_0,U)$. If
    there is only one independent variable, then $T_k = \tau_k^2$ and
    $U = u^2$. The new value of $\mu_k$ is simulated as
    \begin{eqnarray}
      \mu_k|\xi,G,T_k,\mu_0,U & \sim & N_p(\hat{\mu}_k, \Sigma_{\hat{\mu}_k}) 
         \label{eq-mucond1} \\
      \hat{\mu}_k & = & (U^{-1} + n_k T_k^{-1})^{-1} (U^{-1} \mu_0 + n_k T_k^{-1} \bar{\xi}_k)
         \label{eq-gmuhat} \\
      \bar{\xi}_k & = & \frac{1}{n_k} \sum_{i=1}^n G_{ik} \xi_i \label{eq-xibar1} \\
      \Sigma_{\hat{\mu}_k} & = & (U^{-1} + n_k T_k^{-1})^{-1}. \label{eq-gmuhat_sig}
    \end{eqnarray}
  \item
    \label{i-tausqr}
    Draw a new value of $\tau_k^2$ or $T_k$. The distribution of
    $\tau^2|\xi,\mu$ or $T_k|\xi,\mu$ is derived in a manner similar
    to $\sigma^2|\xi,\eta,\alpha,\beta$, and noting that the prior is
    conjugate for this likelihood. The distribution of
    $\tau_k^2|\xi,\mu$ is a scaled inverse-$\chi^2$ distribution, and
    the distribution of $T_k|\xi,\mu$ is an inverse-Wishart
    distribution:
    \begin{enumerate}
    \item
      \label{i-tsqr}
      If there is only one independent variable then draw
      \begin{eqnarray}
	\tau_k^2|\xi,G,\mu_k,w^2 & \sim & \mbox{\rm Inv-}\chi^2(\nu_k, t_k^2) \label{eq-taucond} \\
	\nu_k & = & n_k + 1 \\
	t_k^2 & = & \frac{1}{n_k+1}\left[w^2 + \sum_{i=1}^n G_{ik}(\xi_i - \mu_k)^2 \right]. 
	\label{eq-tsqr}
      \end{eqnarray}
    \item
      \label{i-tcovar}
      If there are multiple independent variables then draw
      \begin{eqnarray}
	T_k|\xi,G,\mu_k,W & \sim & \mbox{\rm Inv-Wishart}_{\nu_k}(S_k) \label{eq-covcond} \\
	\nu_k & = & n_k + p \label{eq-mnu} \\
	S_k & = & W + \sum_{i=1}^n G_{ik}(\xi_i - \mu_k) (\xi_i - \mu_k)^T. \label{eq-smat}
      \end{eqnarray}
    \end{enumerate}
  \item
    \label{i-mu0}
    Draw a new value for $\mu_0|\mu,U$. Noting that conditional on
    $\mu_0$ and $U$, $\mu_1,\ldots,\mu_K$ are independently
    distributed as $N_p(\mu_0,U)$, it is straight-forward to show that
    \begin{eqnarray}
      \mu_0|\mu,U & \sim & N_p(\bar{\mu},U / K) \label{eq-mu0} \\
      \bar{\mu} & = & \frac{1}{K} \sum_{k=1}^K \mu_k. \label{eq-mubar}
    \end{eqnarray}
    If there is only one covariate then $p=1$ and $U = u^2$.
  \item
    \label{i-usqr0}
    Draw a new value for $u^2$ or $U$, given $\mu_0,\mu,$ and $w^2$
    (or $W$). Similar to the case for $\tau^2_k$ or $T_k$, the
    conditional distribution of $u^2$ or $U$ is scaled
    inverse-$\chi^2$ or inverse-Wishart.
    \begin{enumerate}
    \item
      \label{i-usqr}
      If there is only one covariate then
      \begin{eqnarray}
	u^2|\mu_0,\mu,w^2 & \sim & \mbox{\rm Inv-}\chi^2(\nu_u,\hat{u}^2) \label{eq-usqr} \\
	\nu_u & = & K + 1 \label{eq-nuu} \\
	\hat{u}^2 & = & \frac{1}{\nu_u} \left[w^2 + \sum_{k=1}^K (\mu_k - \mu_0)^2 \right].
	\label{eq-usqrhat}
      \end{eqnarray}
    \item
      \label{i-umat}
      If there are multiple covariates then
      \begin{eqnarray}
	U|\mu_0,\mu,W & \sim & \mbox{\rm Inv-Wishart}_{\nu_U}(\hat{U}) \label{eq-umat} \\
	\nu_U & = & K + p \label{eq-nuucap} \\
	\hat{U} & = & W + \sum_{k=1}^K (\mu_k - \mu_0)(\mu_k - \mu_0)^T. \label{eq-uhatmat}
      \end{eqnarray}
    \end{enumerate}
  \item
    \label{i-wsqr0}
    Finally, draw a new value of $w^2|u^2,\tau^2$ or $W|U,T$:
    \begin{enumerate}
    \item
      \label{i-wsqr}
      If there is only one covariate then $w^2|u^2,\tau^2$ is drawn
      from a Gamma distribution. This can be derived by noting that
      $p(w^2|u^2,\tau^2) \propto p(u^2|w^2) p(\tau^2|w^2)$ has the
      form of a Gamma distribution as a function of $w^2$. The new
      value of $w^2$ is then simulated as
      \begin{eqnarray}
	w^2|u^2,\tau^2 & \sim & \mbox{\rm Gamma}(a,b) \label{eq-wsqr} \\
	a & = & \frac{1}{2}(K + 3) \label{eq-a} \\
	b & = & \frac{1}{2} \left[ \frac{1}{u^2} + \sum_{k=1}^K \frac{1}{\tau_k^2} \right].
	\label{eq-b}
      \end{eqnarray}
    \item
      \label{i-wmat}
      If there are multiple covariates then $W|U,T$ is drawn from a
      Wishart distribution. This can be derived by noting that
      $p(W|U,T) \propto p(U|W)p(T|W)$ has the form of a Wishart
      distribution as a function of $W$. The new value of $W$ is then
      simulated as
      \begin{eqnarray}
	W|U,T & \sim & \mbox{\rm Wishart}_{\nu_W}(\hat{W}) \label{eq-wmat} \\
	\nu_W & = & (K + 2)p + 1 \label{eq-nuw} \\
	\hat{W} & = & (U^{-1} + \sum_{k=1}^K T_k^{-1})^{-1}. \label{eq-wmathat}
      \end{eqnarray}
    \end{enumerate}
  \end{enumerate}

  After completing steps \ref{i-ycens}--\ref{i-wsqr0} above, an
  iteration of the Gibbs sampler is complete. One then uses the new
  simulated values of $\xi,\eta,\theta,\psi,$ and the prior
  parameters, and repeats steps \ref{i-ycens}--\ref{i-wsqr0}. The
  algorithm is repeated until convergence, and the values of $\theta$
  and $\psi$ at each iteration are saved. Upon reaching convergence,
  one discards the values of $\theta$ and $\psi$ from the beginning of
  the simulation, and the remaining values of $\alpha, \beta,
  \sigma^2, \mu,$ and $\tau^2$ (or $T$) may be treated as a random
  draw from the posterior distribution, $p(\theta,\psi|x,y)$. One can
  then use these values to calculate estimates of the parameters, and
  their corresponding variances and confidence intervals. The
  posterior distribution of the parameters can also be estimated from
  these values of $\theta$ and $\psi$ using histogram
  techniques. Techniques for monitering convergence of the Markov
  Chains can be found in \citet{gelman04}.

  The output from the Gibbs sampler may be used to perform Bayesian
  inference on other quantities of interest. In particular, the
  Pearson linear correlation coefficient, $\rho$, is often used in
  assessing the strength of a relationship between the $x$ and $y$. A
  random draw from the posterior distribution for the correlation
  between $\eta$ and $\xi_j$, denoted as $\rho_j$, can be calculated
  from Equation (\ref{eq-rho}) for each draw from the Gibbs
  sampler. For the Gaussian mixture model, the variance $Var(\eta)$
  and covariance matrix $\Sigma_{\xi} \equiv Var(\xi)$ are
  \begin{eqnarray}
    Var(\eta) & = & \beta^T \Sigma_{\xi} \beta + \sigma^2 \label{eq-etavar} \\
    \Sigma_{\xi} & = & \sum_{k=1}^K \pi_k (T_k + \mu_k \mu_k^T) - \bar{\xi} \bar{\xi}^T 
    \label{eq-xicovar} \\
    \bar{\xi} & = & \sum_{k=1}^K \pi_k \mu_k \label{eq-xibar},
  \end{eqnarray}
  and $Var(\xi_j)$ is the $j^{\rm th}$ diagonal element of
  $\Sigma_{\xi}$. The simplification for one covariate is
  self-evident.

  If there is considerable posterior probability near $\sigma^2
  \approx 0$ or $\tau_k^2 \approx 0$, then the Gibbs sampler can get
  `stuck'. For example, if $\tau_k^2 \approx 0$, then step \ref{i-xi1}
  of the Gibbs sampler will draw values of $\xi|G \approx
  \mu_k$. Then, step \ref{i-mu} will produce a new value of $\mu_k$
  that is almost identical to the previous iteration, step
  \ref{i-tsqr} will produce a new value of $\tau_k^2 \approx 0$, and
  so on. The Gibbs sampler will eventually get `unstuck', but this can
  take a long time and result in very slow convergence. In particular,
  it is very easy for the Gibbs sampler to get stuck if the
  measurement errors are large relative to $\sigma^2$ or $\tau_k^2$,
  or if the number of data points is small. In this situation I have
  found it useful to use the Metropolis-Hastings algorithm instead.

  \subsubsection{Metropolis-Hastings Algorithm}

  \label{s-metro}

  If the selection function is not independent of $y$, given the
  independent variables (cf. Eq.[\ref{eq-trunclik}]), or if the
  selection function depends on $x$ and the measurement errors are
  correlated, then posterior simulation based on the Gibbs sampler is
  more complicated. In addition, if the measurement errors are large
  compared to the intrinsic dispersion in the data, or if the sample
  size is small, then the Gibbs sampler can become stuck and extremely
  inefficient. In both of these cases one can use the
  Metropolis-Hastings algorithm \citep{metro49,metro53,hast70} to
  sample from the posterior distribution, as the Metropolis-Hasting
  algorithm can avoid constructing markov chains for $\xi$ and
  $\eta$. For a description of the Metropolis-Hastings algorithm, we
  refer the reader to \citet{chib95} or \citet{gelman04}.

  \section{SIMULATIONS}

  \label{s-sims}

  In this section I perform simulations to illustrate the
  effectiveness of the Gaussian structural model for estimating the
  regression parameters, even in the presence of severe measurement
  error and censoring. In addition, I compare the OLS, BCES($Y|X$),
  and FITEXY estimators with a maximum-likelihood estimator based on
  the Gaussian mixture model with $K = 1$ Gaussian.

  \subsection{Simulation Without Non-Detections}

  \label{s-sims1}

\begin{figure}
    \begin{center}
      \scalebox{0.5}{\rotatebox{90}{\plotone{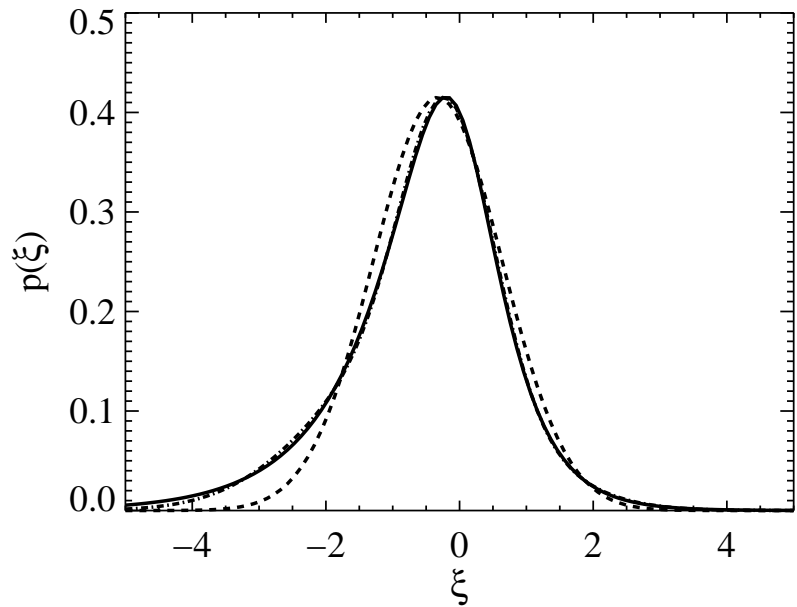}}}
      \caption{The actual distribution of $\xi$ (solid line) for the
      simulations, compared with the best-fitting one (dashed line)
      and two (dashed-dotted line) Gaussian fit. The two Gaussian fit
      is nearly indistinguishable from the true $p(\xi)$. Althought
      the one Gaussian fit provides a reasonable approximation to the
      distribution of $\xi$, it is not able to pick up the asymmetry
      in $p(\xi)$.\label{f-xidist}}
    \end{center}
  \end{figure}

\begin{figure}
    \begin{center}
      \scalebox{0.5}{\rotatebox{90}{\plotone{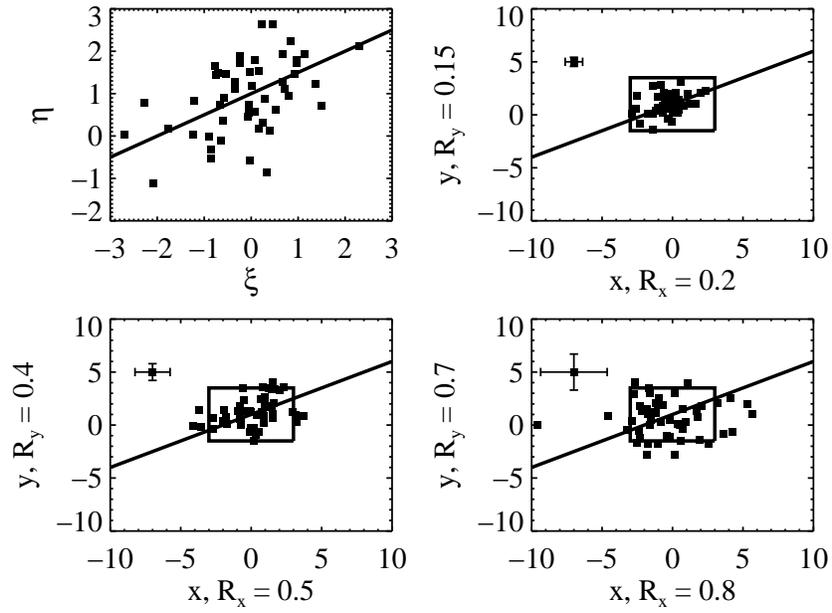}}}
      \caption{Distributions of the simulated data for various levels
      of measurement error (cf., \S~\ref{s-sims1}). The top left panel
      shows the distribution of $\eta$ as a function of $\xi$ for one
      simulated data set; the solid line is the true value of the
      regression line. The remaining panels show the distributions of
      the observed values, $y$ and $x$, for various levels of
      measurement error. The data point with error bars in each panel
      is a fictitious data point and is used to illustrate the median
      values of the error bars. The box outlines the bounds of the
      plot of $\eta$ against $\xi$. As can be seen, large measurement
      errors wash out any visual evidence for a correlation between
      the variables.\label{f-simdist1}}
    \end{center}
  \end{figure}

  The first simulation I performed is for a simple regression with
  one independent variable. I generated $2.7 \times 10^5$ data sets
  by first drawing $n$ values of the independent variable, $\xi$, from
  a distribution of the form
  \begin{equation}
    p(\xi) \propto e^{\xi} \left( 1 + e^{2.75\xi} \right)^{-1}.
    \label{eq-xidistsim}
  \end{equation}
  The distribution of $\xi$ is shown in Figure \ref{f-xidist}, along
  with the best-fitting one and two Gaussian approximations. In this
  case the two Gaussian mixture is nearly indistinguishable from the
  actual distribution of $\xi$, and thus should provide an excellent
  approximation to $p(\xi)$. The values for $\xi$ had a mean of $\mu =
  -0.493$ and a dispersion of $\tau = 1.200$. I varied the number of
  data points in the simulated data sets as $n = 25, 50,$ and
  $100$. I then simulated values of $\eta$ according to Equation
  (\ref{eq-adderr}), with $\alpha = 1.0$ and $\beta = 0.5$. The
  intrinsic scatter, $\epsilon$, had a normal distribution with mean
  zero and standard deviation $\sigma = 0.75$, and the correlation
  between $\eta$ and $\xi$ was $\rho \approx 0.62$. The joint
  distribution of $\xi$ and $\eta$ for one simulated data set with $n
  = 50$ is shown in Figure \ref{f-simdist1}.

  Measured values for $\xi$ and $\eta$ were simulated according to
  Equations (\ref{eq-xerr}) and (\ref{eq-yerr}). The measurement
  errors had a zero mean normal distribution of varying dispersion and
  were independent for $x$ and $y$. The variances in the measurement
  errors, $\sigma^2_{x,i}$ and $\sigma^2_{y,i}$, were different for
  each data point and drawn from a scaled inverse-$\chi^2$
  distribution. The degrees of freedom for the inverse-$\chi^2$
  distribution was $\nu = 5$, and the scale parameters are denoted as
  $t$ and $s$ for the $x$ and $y$ measurement error variances,
  respectively. The scale parameters dictate the typical size of the
  measurements errors, and were varied as $t = 0.5\tau, \tau, 2\tau$
  and $s = 0.5\sigma, \sigma, 2\sigma$. These values corresponded to
  values of $R_x \sim 0.2, 0.5, 0.8$ and $R_y \sim 0.15, 0.4, 0.6$
  respectively. I simulated $10^4$ data sets for each grid point of
  $t, s,$ and $n$, giving a total of $2.7 \times 10^5$ simulated data
  sets. The joint distributions of $x$ and $y$ for varying values of
  $t / \tau$ and $s / \sigma$ are also shown in Figure
  \ref{f-simdist1}. These values of $x$ and $y$ are the `measured'
  values of the simulated data set shown in the plot of $\eta$ as a
  function of $\xi$.

\begin{figure}
    \begin{center}
      \scalebox{0.5}{\rotatebox{90}{\plotone{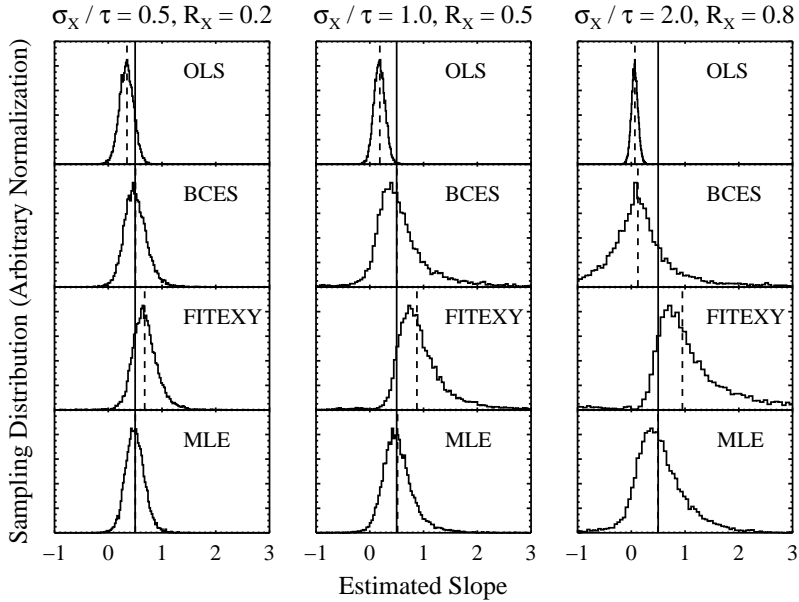}}}
      \caption{The sampling distributions of the slope estimators as a
      function of covariate measurement error magnitude for $n = 50$
      data points and $\sigma_y \sim \sigma$, inferred from
      simulations (cf., \S~\ref{s-sims1}). The estimators are the
      ordinary least-squares estimator (OLS), the BCES($Y|X$)
      estimator, the FITEXY estimator, and the maximum-likelihood
      estimator (MLE) of the $K = 1$ gaussian structural model. The
      solid vertical lines mark the true value of $\beta = 0.5$, and
      the dashed vertical lines mark the median values of each
      respective estimator. The OLS estimator is biased toward zero,
      while the FITEXY estimator is biased away from zero; in both
      cases, the bias gets worse for larger measurement errors. The
      BCES($Y|X$) estimator is, in general, unbiased, but can become
      biased and highly variable if the measurement errors becomes
      large. The MLE of the Gaussian model performs better than the
      other estimators, as it is approximately unbiased and less
      variable. \label{f-sampdist1}}
    \end{center}
  \end{figure}

\begin{figure}
    \begin{center}
      \scalebox{0.5}{\rotatebox{90}{\plotone{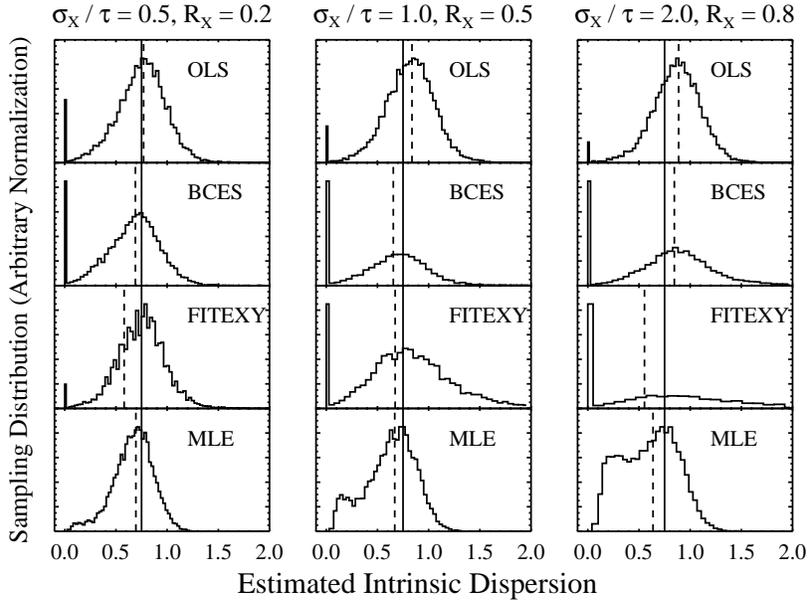}}}
      \caption{Same as Figure \ref{f-sampdist1}, but for the standard
      deviation of the intrinsic scatter, $\sigma$. The solid vertical
      lines mark the true value of $\sigma = 0.75$, and the dashed
      vertical lines mark the median values of each respective
      estimator. All of the estimators exhibit some bias, and the BCES
      and FITEXY estimators can exhibit significant
      variance. Moreover, the BCES and FITEXY estimators both commonly
      have values of $\hat{\sigma} = 0$, misleading one into
      concluding that there is no intrinsic scatter; this occasionally
      occurs for the OLS estimate as well. In contrast, the MLE based
      on the Gaussian model does not suffer from this problem, at
      least for these simulations. \label{f-sampdist2}}
    \end{center}
  \end{figure}

  For each simulated data set, I calculated the maximum-likelihood
  estimate, found by maximizing Equation (\ref{eq-bobslik}). For
  simplicity, I only use $K = 1$ Gaussian. I also calculated the
  OLS, BCES($Y|X$), and FITEXY estimates for comparison. I calculated
  a OLS estimate of $\sigma^2$ by subtracting the average $\sigma^2_y$
  from the variance in the regression residuals. If the OLS estimate
  of $\sigma^2$ was negative, I set $\hat{\sigma}_{OLS} =
  0$. Following \citet{fuller87}, I estimate $\sigma^2$ for a
  BCES($Y|X$)-type estimator as $\hat{\sigma}^2_{BCES} = Var(y) -
  \bar{\sigma}^2_y - \hat{\beta}_{\rm BCES} Cov(x,y)$, where
  $\bar{\sigma}^2_y$ is the average measurement error variance in $y$,
  and $\hat{\beta}_{\rm BCES}$ is the BCES($Y|X$) estimate of the
  slope. If $\hat{\sigma}^2_{BCES}$ is negative, I set
  $\hat{\sigma}_{BCES} = 0$. Following \citet{trem02}, I compute a
  FITEXY estimate of $\sigma$ by increasing $\sigma^2$ until
  $\chi^2_{EXY} / (n - 2) = 1$, or assume $\sigma^2 = 0$ if
  $\chi^2_{EXY} / (n - 2) < 1$. The sampling distributions of the
  slope and intrinsic scatter estimators for $n = 50$ are shown in
  Figures \ref{f-sampdist1} and \ref{f-sampdist2} as a function of $t
  / \tau$, and the results of the simulations are summarized in Table
  \ref{t-univest}.

  The bias of the OLS estimate is apparent, becoming more severe as
  the measurement errors in the independent variable increase. In
  addition, the variance in the OLS slope estimate decreases as the
  measurement errors in $\xi$ increase, giving one the false
  impression that one's estimate of the slope is more precise when the
  measurement errors are large. This has the effect of concentrating
  the OLS estimate of $\beta$ around $\hat{\beta}_{OLS} \sim 0$, thus
  effectively erasing any evidence of a relationship between the two
  variables. When the measurement errors are large, the OLS estimate
  of the intrinsic scatter, $\hat{\sigma}^2_{OLS}$, is occasionally
  zero.

  The BCES($Y|X$) estimator performs better than the OLS and FITEXY
  estimators, being approximately unbiased when the measurement errors
  are $\sigma_x / \tau \lesssim 1$. However, the BCES estimate of the
  slope, $\hat{\beta}_{\rm BCES} = Cov(x,y) / (Var(x) -
  \bar{\sigma}^2_x)$, suffers some bias when the measurement errors
  are large and/or the sample size is small. In addition, the variance
  in $\hat{\beta}_{BCES}$ is larger than the MLE, and
  $\hat{\beta}_{BCES}$ becomes considerably unstable when the
  measurement errors on $\xi$ are large. This instability results
  because the denominator in the equation for $\hat{\beta}_{\rm BCES}$
  is $Var(x) - \bar{\sigma}^2_x$. If $\bar{\sigma}^2_x \approx
  Var(x)$, then the denominator is $\approx 0$, and $\hat{\beta}_{\rm
  BCES}$ can become very large. Similar to the OLS and FITEXY
  estimates, the estimate of the intrinsic variance for the BCES-type
  estimator is often zero when the measurement errors are large,
  suggesting the false conclusion that there is no intrinsic scatter
  about the regression line.

  The FITEXY estimator performed poorly in the simulations, being both
  biased and highly variable. The bias of the FITEXY estimator is such
  that $\hat{\beta}_{EXY}$ tends to overestimate $\beta$, the severity
  of which tends to increase as $R_y$ decreases. This upward bias in
  $\hat{\beta}_{EXY}$ has been noted by \citet{weiner06}, who also
  performed simulations comparing $\hat{\beta}_{EXY}$ with
  $\hat{\beta}_{BCES}$. They note that when one minimizes
  $\chi^2_{EXY}$ alternatively with respect to $\beta$ and $\sigma^2$,
  and iterates until convergence, then the bias in $\hat{\beta}_{EXY}$
  can be improved. I have tested this and also find that the bias in
  $\hat{\beta}_{EXY}$ is reduced, but at the cost of a considerable
  increase in variance in $\hat{\beta}_{EXY}$. In general, our
  simulations imply that the variance of the FITEXY estimator is
  comparable to that of the BCES($Y|X$) estimator if one does not
  iterate the minimization of $\chi^2_{EXY}$, and the variance of
  $\hat{\beta}_{EXY}$ is larger if one does iterate. However, since
  $\hat{\beta}_{BCES}$ is approximately unbiased when $R_x$ is not too
  large, $\hat{\beta}_{BCES}$ should be preferred over
  $\hat{\beta}_{EXY}$. In addition, when the measurement errors are
  large the FITEXY estimate of $\sigma$ is commonly
  $\hat{\sigma}_{EXY} = 0$, similar to the BCES-type estimate of the
  intrinsic dispersion.

  The maximum-likelihood estimator based on the Gaussian structural
  model performs better than the OLS, BCES, and FITEXY estimators, and
  gives fairly consistent estimates even in the presence of severe
  measurement error and low sample size. The MLE is approximately
  unbiased, in spite of the fact that the MLE incorrectly assumes that
  the independent variables are normally distributed. The variance in
  the MLE of the slope, $\hat{\beta}_{MLE}$, is smaller than that of
  $\hat{\beta}_{BCES}$ and $\hat{\beta}_{EXY}$, particularly when
  $R_x$ is large. In contrast to the OLS estimate of the slope, the
  dispersion in $\hat{\beta}_{MLE}$ increases as the measurement
  errors increases, reflecting the additional uncertainty in
  $\hat{\beta}_{MLE}$ caused by the measurement errors. Finally, in
  contrast to the other estimators, the MLE of the intrinsic variance
  is always positive, and the probability of obtaining
  $\hat{\sigma}_{MLE} = 0$ is negligible for these simulations.

  I argued in \S~\ref{s-normreg} that assuming a uniform distribution
  on $\xi$ does not lead to better estimates than the usual OLS
  case. I also used these simulations to estimate the sampling
  density of the MLE assuming $p(\xi) \propto 1$. The results were
  nearly indistinguishable from the OLS estimator, supporting our
  conjecture that assuming $p(\xi) \propto 1$ does not offer an
  improvement over OLS.

\begin{figure}
    \begin{center}
      \scalebox{0.5}{\rotatebox{90}{\plotone{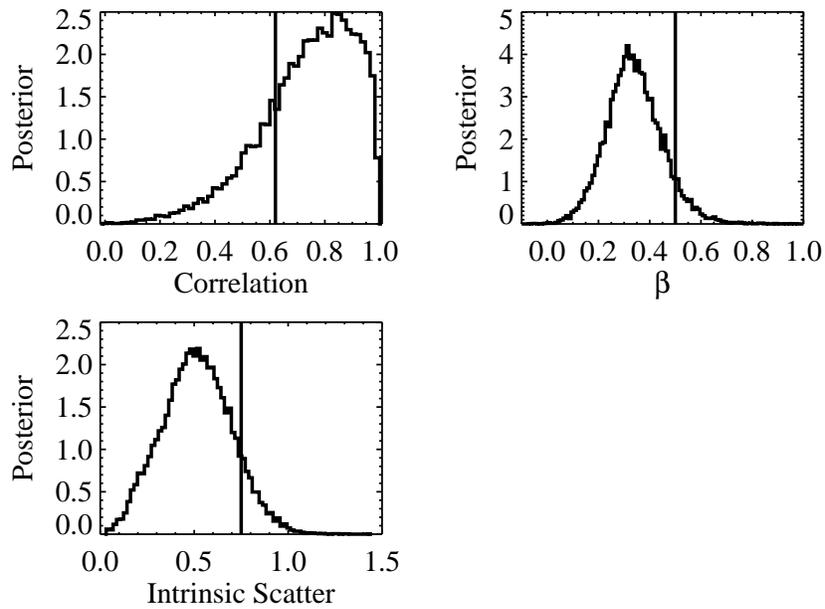}}}
      \caption{The marginal posterior distributions of the linear
      correlation coefficient, the regression slope, and the intrinsic
      dispersion for a simulated data set of $n = 50$ data points with
      $\sigma_x \sim \tau$ and $\sigma_y \sim \sigma$. The vertical
      lines mark the true values of the parameters. The true values of
      the regression parameters are contained within the spread of the
      marginal posteriors, implying that bounds on the regression
      parameters inferred from the posterior are
      trustworthy.\label{f-post1d}}
    \end{center}
  \end{figure}

  While it is informative to compare the sampling distribution of our
  proposed maximum-likelihood estimator with those of the OLS,
  BCES($Y|X$), and FITEXY estimators, I do not derive the
  uncertainties in the regression parameters from the sampling
  distribution of the MLE. As described in \S~\ref{s-markov}, we
  derive the uncertainties in the regression parameters by simulating
  draws from the posterior distribution, $p(\theta,\psi|x,y)$. This
  allows a straight-forward method of interpreting the parameter
  uncertainties that does not rely on large-sample approximations, as
  the posterior distribution is the probability distribution of the
  parameters, given the observed data. The posterior distributions of
  $\rho, \beta,$ and $\sigma$ for a simulated data set with $n = 50,
  \sigma_x \sim \tau,$ and $\sigma_y \sim \sigma$ is shown in Figure
  \ref{f-post1d}. When estimating these posteriors, I used $K = 2$
  Gaussians in the mixture model. As can be seen from Figure
  \ref{f-post1d}, the true values of $\rho, \beta,$ and $\sigma$ are
  contained within the regions of non-negligible posterior
  probability. I have estimated posteriors for other simulated data
  sets, varying the number of data points and the degree of
  measurement error. As one would expect, the uncertainties in the
  regression parameters, represented by the widths of the posterior
  distributions, increase as the size of the measurement errors
  increase and the sample size decreases.

\begin{figure}
    \begin{center}
      \scalebox{0.5}{\rotatebox{90}{\plotone{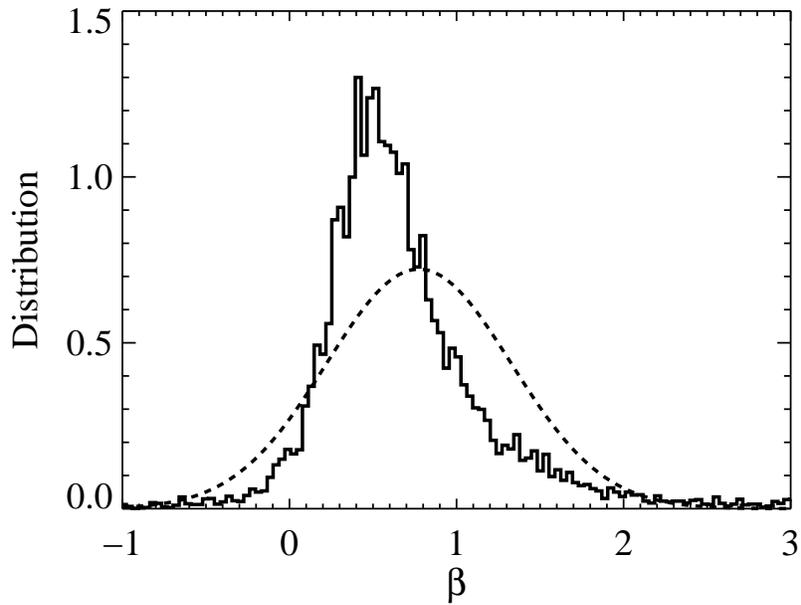}}}
      \caption{The posterior distributions of the slope (solid
      histogram), compared with the posterior approximated from the
      MLE and Fisher information matrix (dashed line), for a simulated
      data set of $n = 50$ data points with $\beta = 0.5, \sigma_x
      \sim \tau,$ and $\sigma_y \sim \sigma$. The two distributions
      have not converged and the bayesian and frequentist inference
      differ in this case, with the bayesian approach placing more
      probability near $\beta \approx 0.5$ and on positive value of
      $\beta$. \label{f-fbcompare}}
    \end{center}
  \end{figure}

  A common frequentist approach is to compute the covariance matrix of
  the MLE by inverting the estimated Fisher information matrix,
  evaluated at the MLE. Then, under certain regularity conditions, the
  MLE of the parameters is asympotically normally distributed with
  mean equal to the true value of the parameters and covariance matrix
  equal to the inverse of the Fisher information matrix. Furthermore,
  under these regularity conditions the posterior distribution and
  sampling distribution of the MLE are asymptotically the same. Figure
  \ref{f-fbcompare} compares the posterior distribution of the slope
  for a simulated data set with that inferred from the MLE. The
  posterior and MLE was calculated assuming $K = 1$ Gaussian. As can
  be seen, the posterior distribution for $\beta$ is considerably
  different from the approximation based on the MLE of $\beta$, and
  thus the two have not converged for this sample. In particular, the
  posterior is more skewed and heavy-tailed, placing more probability
  on values of $\beta > 0$ than does the distribution approximated by
  the MLE. Therefore, uncertainties in the MLE should be interpreted
  with caution if using the asymptotic approximation to the sampling
  distribution of the MLE.

  \subsection{Simulation With Non-Detections}

  \label{s-sims2}

  To assess the effectiveness of the Gaussian structural model in
  dealing with censored data sets with measurement error, I introduced
  non-detections into the simulations. The simulations were performed
  in an identical manner as that described in \S~\ref{s-sims1}, but
  now I only consider sources to be `detected' if $y > 1.5$. For those
  sources that were `censored' ($y < 1.5$), I placed an upper limit on
  them of $y = 1.5$.
  
\begin{figure}
    \begin{center}
      \includegraphics[scale=0.33,angle=90]{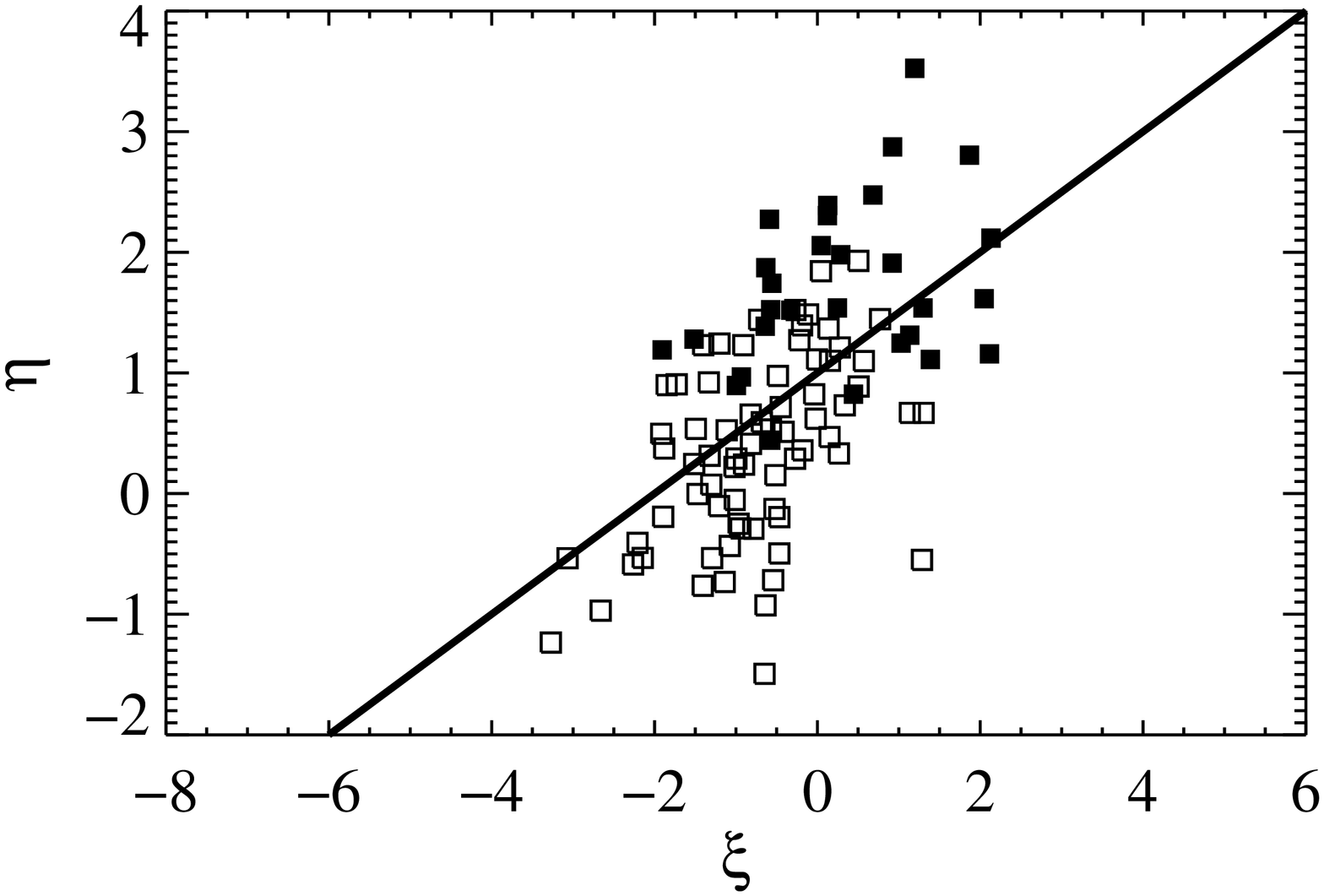}
      \includegraphics[scale=0.33,angle=90]{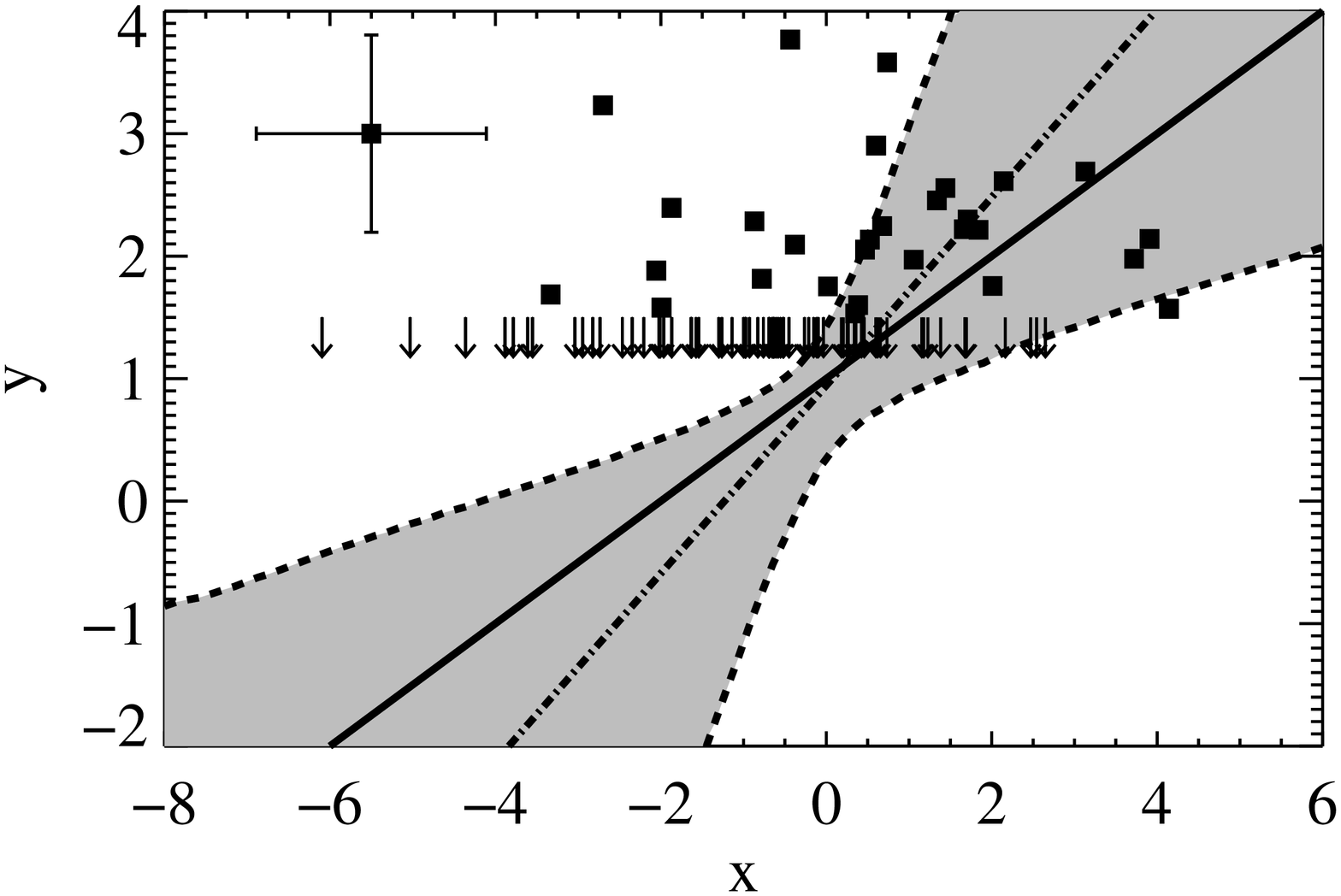}
      \caption{Distribution of $\eta$ and $\xi$ (left), and the
      measured values of $y$ and $x$ (right), from a simulated
      censored data set of $n = 50$ data points, $\sigma_x \sim \tau$,
      and $\sigma_y \sim \sigma$ (cf., \S~\ref{s-sims2}). In the plot
      of $\eta$ and $\xi$, the solid squares denote the values of
      $\xi$ and $\eta$ for the detected data points, and the hollow
      squares denote the values of $\xi$ and $\eta$ for the undetected
      data points. The solid line in both plots is the true regression
      line. In the plot of $y$ and $x$, the squares denote the
      measured values of $x$ and $y$ for the detected data points, and
      the arrows denote the `upper limits' on $y$ for the undetected
      data points. The fictitious data point with error bars
      illustrates the median values of the error bars. The
      dashed-dotted line shows the best fit regression line, as
      calculated from the posterior median of $\alpha$ and $\beta$,
      and the filled region defines the approximate $95\%$ $(2\sigma)$
      pointwise confidence intervals on the regression line. The true
      values of the regression line are contained within the $95\%$
      confidence intervals.\label{f-cens_reg}}.
    \end{center}
  \end{figure}

  I focus on the results for a simulated data set with $n = 100$ data
  points and measurement errors similar to the intrinsic dispersion in
  the data, $\sigma_y \sim \sigma$ and $\sigma_x \sim \tau$. The
  detection threshold of $y > 1.5$ resulted in a detection fraction of
  $\sim 30\%$. This simulation represents a rather extreme case of
  large measurement errors and low detection fraction, and provides an
  interesting test of the method. In Figure \ref{f-cens_reg} I show
  the distribution of $\xi$ and $\eta$, as well as the distribution of
  their measured values, for one of the simulated data sets. For this
  particular data set, there were 29 detections and 71
  non-detections. As can be seen, the significant censoring and large
  measurement errors have effectively erased any visual evidence for a
  relationship between the two variables.

\begin{figure}
    \begin{center}
      \scalebox{0.5}{\rotatebox{90}{\plotone{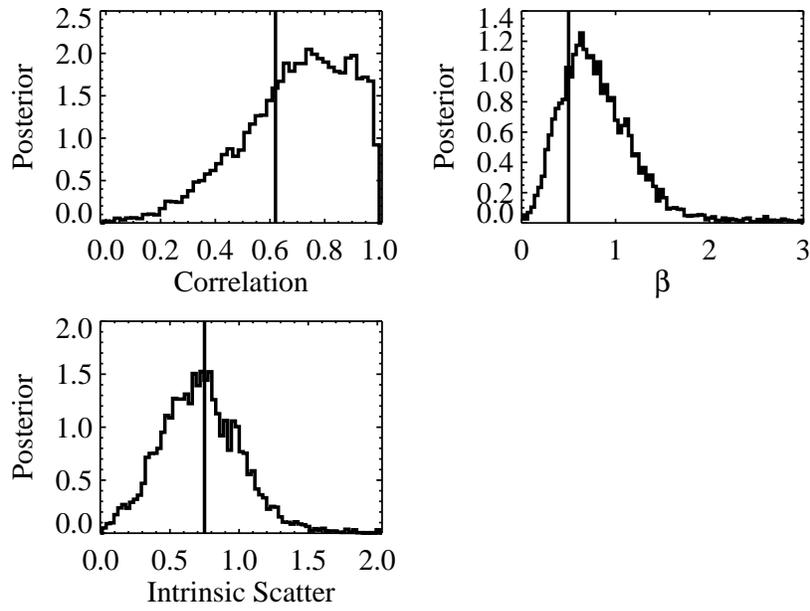}}}
      \caption{Same as Figure \ref{f-post1d}, but for the censored
      data set shown in Figure \ref{f-cens_reg}. The true values of
      the regression parameters are contained within the spread of the
      posteriors, implying that bounds on the regression parameters
      inferred from the posterior are trustworthy. \label{f-censpost}}
    \end{center}
  \end{figure}

  I estimated the posterior distribution of the regression parameters
  for this data set using the Gibbs sampler (cf, \S~\ref{s-gibbs})
  with $K = 2$ Gaussians. The posterior median of the regression line,
  as well as the $95\%\ (2\sigma)$ pointwise confidence
  intervals\footnote{Technically, these are called `credibility
  intervals', as I am employing a Bayesian approach. These intervals
  contain $95\%$ of the posterior probability. While the difference
  between confidence intervals and credibility intervals is not purely
  semantical, I do not find the difference to be significant within
  the context of my work, so I use the more familiar term
  `confidence interval'.} on the regression line are shown in Figure
  \ref{f-cens_reg}. The posterior distributions for $\rho, \beta,$ and
  $\sigma$ are shown in Figure \ref{f-censpost}. As can be seen, the
  true value of the parameters is contained within the $95\%$
  probability regions, although the uncertainty is large. For this
  particular data set, we can put limits on the value of the
  correlation coefficient as $0.2 \lesssim \rho \lesssim 1$ and the
  slope as $0 \lesssim \beta \lesssim 2.0$. For comparison, the usual
  maximum-likelihood estimate that ignores the measurement error
  \citep[e.g.,][]{isobe86} concludes $\hat{\beta} = 0.229 \pm
  0.077$. This estimate is biased and differs from the true value of
  $\beta$ at a level of $3.5\sigma$.

  The posterior constraints on the regression parameters are broad,
  reflecting our considerable uncertainty in the slope, but they are
  sufficient for finding a positive correlation between the two
  variables, $\xi$ and $\eta$. Therefore, despite the high level of
  censoring and measurement error in this data set, we would still be
  able to conclude that $\eta$ increases as $\xi$ increases.

  \section{APPLICATION TO REAL ASTRONOMICAL DATA: DEPENDENCE OF $\Gamma_X$ 
    ON $L_{bol} / L_{Edd}$ FOR RADIO-QUIET QUASARS}

  \label{s-data}

  To further illustrate the effectiveness of the method, I apply it
  to a data set drawn from my work on investigating the X-ray
  properties of radio-quiet quasars (RQQs). Recent work has suggested
  a correlation between quasar X-ray spectral slope, $\alpha_X,
  f_{\nu} \propto \nu^{-\alpha_X},$ and quasar Eddington ratio,
  $L_{bol} / L_{Edd}$ \citep[e.g.,][]{porq04,picon05,shemm06}. In this
  section I apply the regression method to a sample of 39 $z < 0.83$
  RQQs and confirm the $\Gamma_X$--$L_{bol} / L_{Edd}$
  correlation. Because the purpose of this section is to illustrate
  the use of this regression method on real astronomical data, I defer
  a more in-depth analysis to a future paper.

  Estimation of the Eddington luminosity, $L_{Edd} \propto M_{BH}$,
  requires an estimate of the black hole mass, $M_{BH}$.  Black hole
  virial masses may be estimated as $M_{BH} \propto R v^2$, where $R$
  is the broad line region size, and $v$ is the velocity dispersion of
  the gas emitting the broad emission lines. A correlation has been
  found between the luminosity of a source and the size of it's broad
  line region \citep[the $R$--$L$ relationship, e.g.,][]{kaspi05}. One
  can then exploit this relationship, and use the broad line $FWHM$ as
  an estimate for $v$, obtaining virial mass estimates $\hat{M}_{BH}
  \propto L^{\theta} v^2$ \citep[e.g.,][]{wand99}, where the exponent
  is $\theta \approx 0.5$ \citep[e.g.,][]{vest06}. Unfortunately, the
  uncertainty on the broad line estimates of $M_{BH}$ can be
  considerable, having a standard deviation of $\sigma_m \sim 0.4$ dex
  \citep[e.g.,][]{bhmmgii,vest06,kelly06a}. For ease of comparison
  with previous work, I estimate $M_{BH}$ using only the H$\beta$
  emission line. The logarithm of the virial mass estimates were
  calculated using the H$\beta$ luminosity and $FWHM$ according to the
  relationship given by \citet{vest06}.

  My sample consists of a subset of the sample of
  \citet{kelly06b}. These sources have measurements of the X-ray
  photon index, $\Gamma_X = \alpha_X + 1$, obtained from
  \emph{Chandra} observations, and measurements of the optical/UV
  luminosity at $2500\AA$, denoted as $L_{2500}$, obtained from SDSS
  spectra. The H$\beta$ profile was modeled as a sum of Gaussians and
  extracted from the SDSS spectra according to the procedure described
  in \citet{kelly06a}. I estimated the H$\beta$ $FWHM$ and luminosity
  from the line profile fits.

\begin{figure}
    \begin{center}
      \includegraphics[scale=0.33,angle=90]{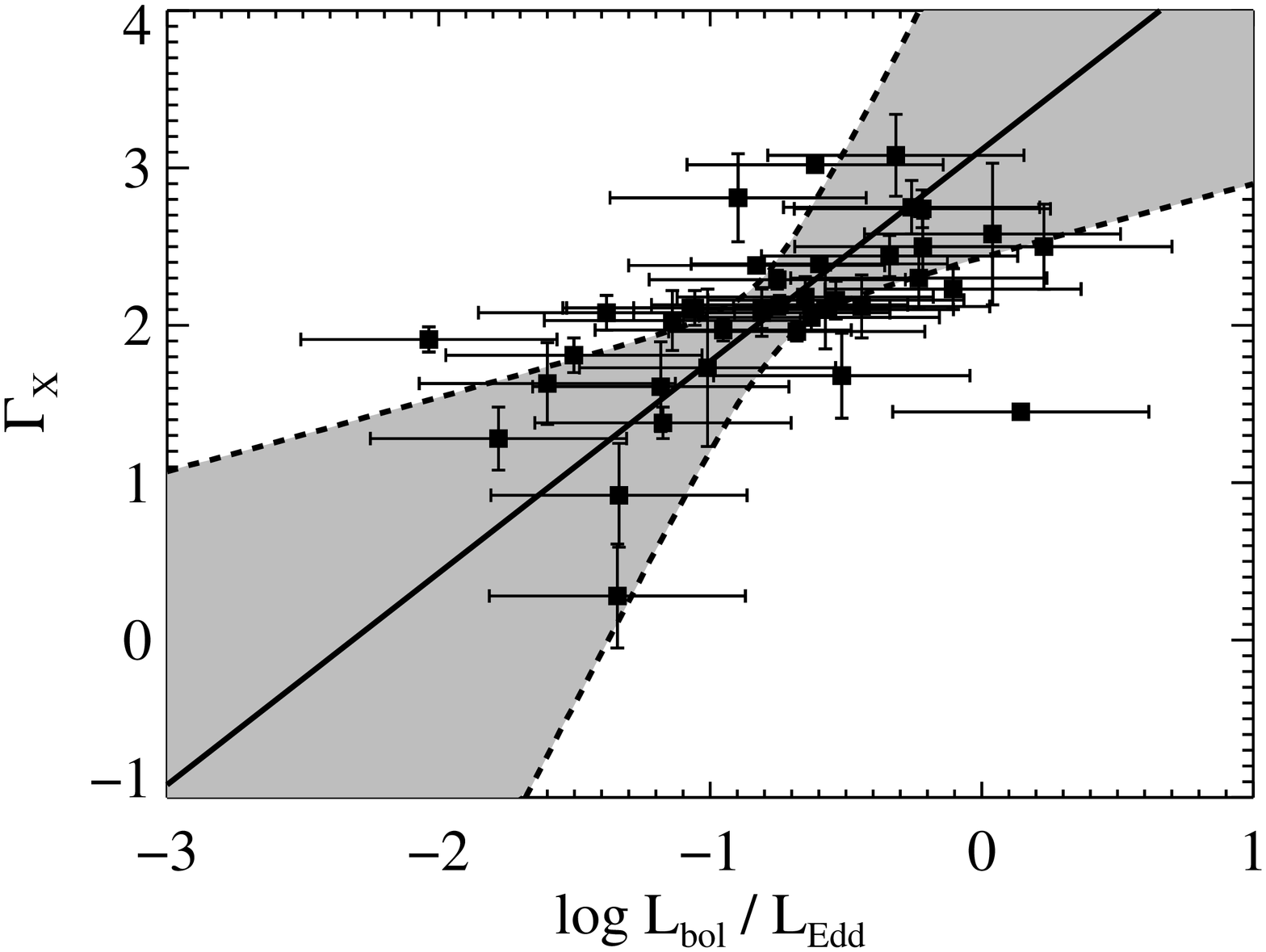}
      \includegraphics[scale=0.33,angle=90]{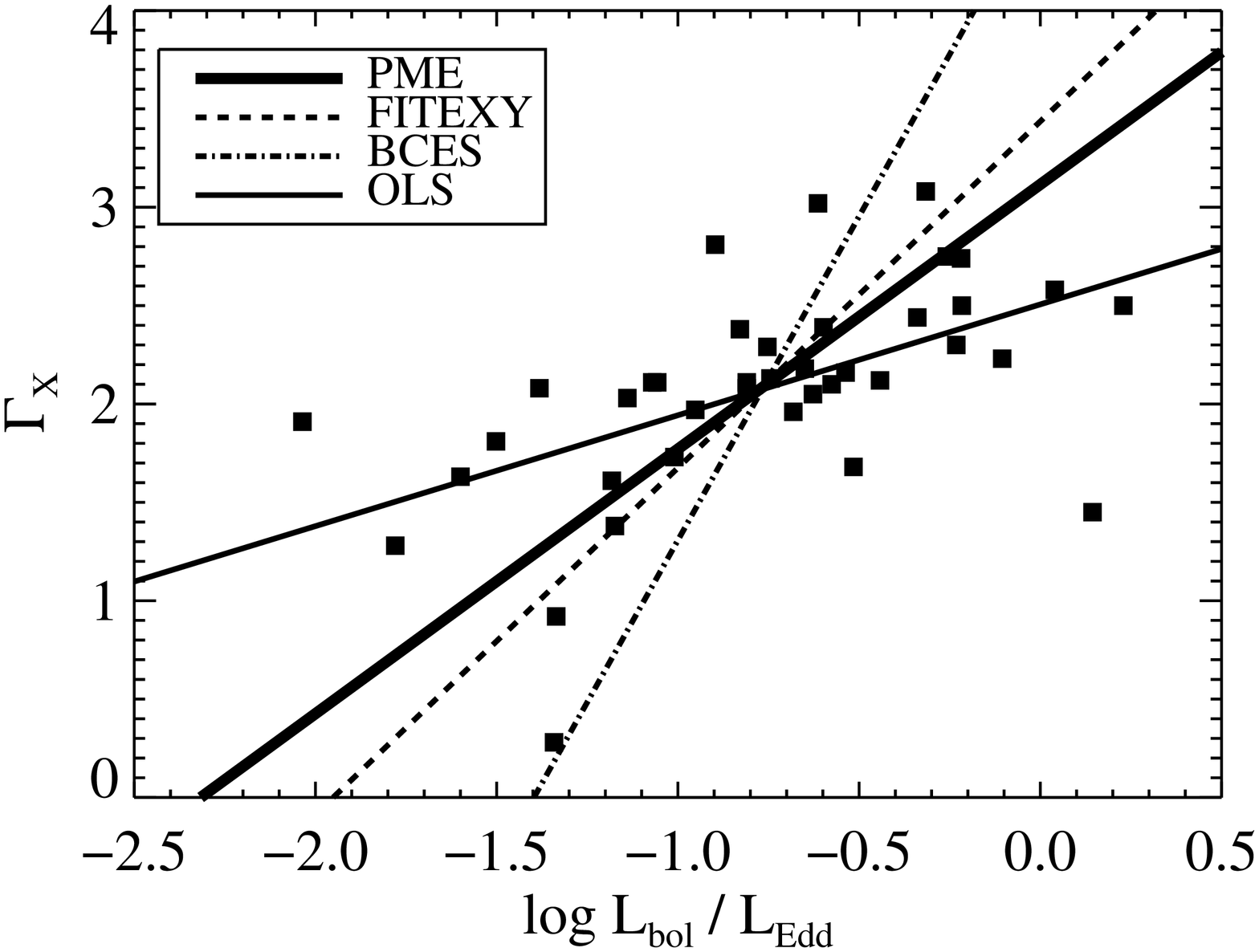}
      \caption{The X-ray photon index, $\Gamma_X$, as a function of
	$\log L_{bol} / L_{Edd}$ for 39 $z \lesssim 0.8$ radio-quiet
	quasars. In both plots the thick solid line shows the
	posterior median estimate (PME) of the regression line. In the
	left plot, the filled region denotes the $95\%$ $(2\sigma)$
	pointwise confidence intervals on the regression line. In the
	right plot, the thin solid line shows the OLS estimate, the
	dashed line shows the FITEXY estimate, and the dot-dashed line
	shows the BCES($Y|X$) estimate; the error bars have been
	omitted for clarity. A significant positive trend is implied
	by the data.\label{f-gamx_eddrat}}
    \end{center}
  \end{figure}

  I estimate the bolometric luminosity, $L_{bol}$, from the
  luminosity at $2500\AA$, assuming a constant bolometric correction
  $L_{bol} = 5.6 L_{2500}$ \citep{elvis94}. The standard deviation in
  this bolometric correction reported by \citet{elvis94} is 3.1,
  implying an uncertainty in $\log L_{bol}$ of $\sigma_{bol} \sim
  0.25$ dex. Combining this with the $\sim 0.4$ dex uncertainty on
  $\log M_{BH}$, the total `measurement error' on $\log L_{bol} /
  L_{Edd}$ becomes $\sigma_x \sim 0.47$ dex. The distribution of
  $\Gamma_X$ as a function of $\log L_{bol} / L_{Edd}$ is shown in
  Figure \ref{f-gamx_eddrat}. As can be seen, the measurement errors
  on both $\Gamma_X$ and $\log L_{bol} / L_{Edd}$ are large and make a
  considerable contribution to the observed scatter in both variables,
  where $R_y \sim 0.1$ and $R_x \sim 0.8$. Therefore, we expect the
  measurement errors to have a significant effect on the correlation
  and regression analysis.

\begin{figure}
    \begin{center}
      \scalebox{0.5}{\rotatebox{90}{\plotone{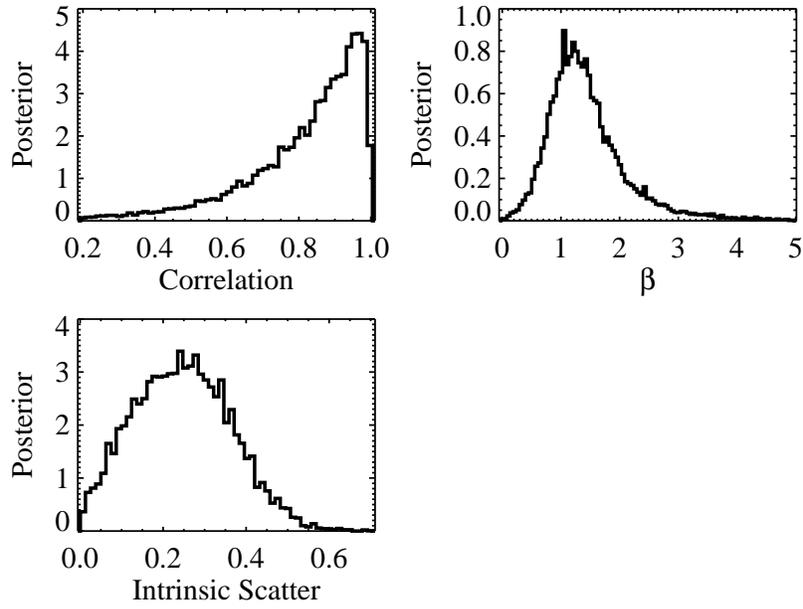}}}
      \caption{Same as Figure \ref{f-post1d}, but for the
      $\Gamma_X$--$\log L_{bol} / L_{Edd}$ regression. Although the
      uncertainty on the slope and correlation are large, the bounds
      on them implied by the data are $0 \lesssim \beta \lesssim 3.5$
      and $0.2 \lesssim \rho \lesssim 1.0$.\label{f-posthb}}
    \end{center}
  \end{figure}

  I performed the regression assuming the linear form $\Gamma_X =
  \alpha + \beta \log L_{bol} / L_{Edd}$, and modelleling the
  intrinsic distribution of $\log L_{bol} / L_{edd}$ using $K = 2$
  Gaussians. Draws from the posterior were obtained using the Gibbs
  sampler. The marginal posterior distributions for $\beta, \sigma$,
  and the correlation between $\Gamma_X$ and $\log L_{bol} / L_{edd}$,
  $\rho$, are shown in Figure \ref{f-posthb}, and the posterior median
  and $95\%$ ($2\sigma$) pointwise intervals on the regression line
  are shown in Figure \ref{f-gamx_eddrat}. The posterior median
  estimate of the parameters are $\hat{\alpha} = 3.12 \pm 0.41$ for
  the constant, $\hat{\beta} = 1.35 \pm 0.54$ for the slope,
  $\hat{\sigma} = 0.26 \pm 0.11$ for the intrinsic scatter about the
  regression line, $\hat{\mu}_{\xi} = -0.77 \pm 0.10$ for the mean of
  $\log L_{bol} / L_{Edd}$, and $\hat{\sigma}_{\xi} = 0.32 \pm 0.12$
  dex for the dispersion in $\log L_{bol} / L_{edd}$. Here, I have
  used a robust estimate of the posterior standard deviation as an
  `error bar' on the parameters. These results imply that the observed
  scatter in $\log L_{bol} / L_{Edd}$ is dominated by measurement
  error, $\sigma_x / \tau \sim 1.5$, as expected from the large value
  of $R_x$.

  For comparison, the BCES($Y|X$) estimate of the slope is
  $\hat{\beta}_{BCES} = 3.29 \pm 3.34$, the FITEXY estimate is
  $\hat{\beta}_{EXY} = 1.76 \pm 0.49$, and the OLS estimate is
  $\hat{\beta}_{OLS} = 0.56 \pm 0.14$; the standard error on
  $\hat{\beta}_{EXY}$ was estimated using bootstrapping. Figure
  \ref{f-gamx_eddrat} also compares the OLS, BCES, and FITEXY best-fit
  lines with the posterior median estimate. The $95\%$ confidence
  region on the slope implied by the posterior draws is $0.46 < \beta
  < 3.44$, whereas the approximate $95\%$ confidence region implied by
  the BCES, FITEXY, and OLS standard errors are $-3.26 < \beta <
  9.84$, $0.80 < \beta < 2.72$, and $0.42 < \beta < 0.70$,
  respectively. The OLS and FITEXY estimates and the Bayesian approach
  give `statistically significant' evidence for a correlation between
  $\log L_{bol} / L_{Edd}$ and $\Gamma_X$; however the BCES estimate
  is too variable to rule out the null hypothesis of no
  correlation. As noted before, the large measurement errors on $\log
  L_{bol} / L_{Edd}$ bias the OLS estimate of $\beta$ toward shallower
  values and the FITEXY estimate of $\beta$ toward steeper
  values. Because of this bias, confidence regions based on
  $\hat{\beta}_{OLS}$ and $\hat{\beta}_{EXY}$ are not valid because
  they are not centered on the true value of $\beta$, and thus do not
  contain the true value with the stated probability (e.g.,
  $95\%$). On the other hand, confidence regions based on the BCES
  estimate are likely to be approximately valid; however, in this
  example the large measurement errors have caused
  $\hat{\beta}_{BCES}$ to be too variable to give meaningful
  constraints on the regression slope.

  The BCES-type estimate of the intrinsic dispersion was
  $\hat{\sigma}_{BCES} = 0.32$ and the OLS estimate of the intrinsic
  dispersion was $\hat{\sigma}_{OLS} = 0.41$, where both were
  calculated in the same manner as in \S~\ref{s-sims1}. The FITEXY
  estimate of the intrinsic dispersion was $\hat{\sigma}_{EXY} = 0$,
  as $\chi^2_{EXY} / (n - 2) < 1$. The BCES-type estimate of $\sigma$
  is similar to the Bayesian posterior median estimate, while
  $\hat{\sigma}_{OLS}$ overestimates the scatter compared to the
  Bayesian estimate by $\approx 58\%$. In contrast, the FITEXY
  estimator does not find any evidence for intrinsic scatter in the
  regression, which is inconsistent with the posterior distribution of
  $\sigma$.

  From the posterior distribution, we can constrain the correlation
  between $\Gamma_X$ and $\log L_{bol} / L_{Edd}$ to be $0.328
  \lesssim \rho \lesssim 0.998$ with $\approx 95\%$ probability,
  confirming the positive correlation between $\Gamma_X$ and Eddington
  ratio seen previously. The posterior median estimate of the
  correlation is $\hat{\rho} = 0.87$, compared with an estimate of
  $\hat{r} = 0.54$ if one naively calculates the correlation directly
  from the measured data. The large measurement errors significantly
  attenuate the observed correlation, making the observed correlation
  between $\Gamma_X$ and $\log L_{bol} / L_{edd}$ appear weaker than
  if one does not correct for the measurement errors.

  \section{CONCLUSIONS}

  \label{s-conclusions}

  In this work I have derived a likelihood function for handling
  measurement errors in linear regression of astronomical data. Our
  probability model assumes that the measurement errors are Gaussian
  with zero mean and known variance, that the intrinsic scatter in the
  dependent variable about the regression line is Gaussian, and that
  the intrinsic distribution of the independent variables can be well
  approximated as a mixture of Gaussians. I extend this model to
  enable the inclusion of non-detections, and describe how to
  incorporate the data selection process. A Gibbs sampler is described
  to enable simulating random draws from the posterior distribution.

  I illustrated the effectiveness of structural Gaussian mixture model
  using simulation. For the specific simulations performed, a
  maximum-likelihood estimator based on the Gaussian structural model
  performed better than the OLS, BCES($Y|X$), and FITEXY estimators,
  especially when the measurement errors were large. In addition, our
  method also performed well when the measurement errors were large
  and the detection fraction was small, with the posterior
  distributions giving reasonable bounds on the regression
  parameters. These results were in spite of the fact that the
  intrinsic distribution of the independent variable was not a sum of
  Gaussians for the simulations, suggesting that approximating the
  distribution of the independent variable as a mixture of Gaussians
  does not lead to a significant bias in the results. Finally, I
  concluded by using the method to fit the radio-quiet quasar X-ray
  photon index as a function of $\log L_{bol} / L_{Edd}$, using a
  sample of 39 $z < 0.83$ sources. The posterior distribution for this
  data set constrained the slope to be $0 \lesssim \beta \lesssim 3.5$
  and the linear correlation coefficient to be $0.2 \lesssim \rho
  \lesssim 1.0$, confirming the correlation between X-ray spectral
  slope and Eddington ratio seen by other authors.

  Although I have focused on linear regression in this work, the
  approach that I have taken is quite general and can be applied to
  other applications. In particular, Equations (\ref{eq-obslik2}),
  (\ref{eq-trunclik}), and (\ref{eq-censlik}) are derived under
  general conditions and are not limited to regression. In this work,
  I assume forms for the respective probability densities that are
  appropriate for linear regression; however, these equation provide a
  framework for constructing more general probability models of one's
  data, as in, for example, nonlinear fitting \citep[e.g.,][]{} or
  estimation of luminosity functions.

  IDL routines for constructing Markov Chains for sampling from the
  posterior are publicly available from the IDL astronomy user's
  library \footnote{\url{http://idlastro.gsfc.nasa.gov/homepage.html}}
  or directly from B.~Kelly.

  \acknowledgements

  This work was supported in part by NSF grant AST-0307384. The author
  would like to thank the referee for comments that contributed to the
  improvement of this paper, and for providing some of the references
  to the statistics literature. The author would also like to thank
  Jill Bechtold, Eric Feigelson, and Aneta Siemiginowska for looking
  over and offering helpful comments on an early version of this paper.

  \clearpage

  \clearpage

  \clearpage

  \clearpage

  \clearpage

  \clearpage

  \clearpage

  \clearpage

  \clearpage

  \clearpage

  \clearpage

\clearpage

\begin{deluxetable}{cccccccccc}
  \tabletypesize{\scriptsize}
  \rotate
  \tablecaption{Dependence of the Estimator Sampling Distributions on 
    Measurement Error and Sample Size\label{t-univest}}
  \tablewidth{0pt}
  \tablehead{
    & & \multicolumn{2}{c}{OLS} & 
    \multicolumn{2}{c}{BCES($Y|X$)} &
    \multicolumn{2}{c}{FITEXY} &
    \multicolumn{2}{c}{MLE} \\
    \colhead{$t / \tau = s / \sigma$\tablenotemark{a}}
    & \colhead{$n$\tablenotemark{b}}
    & \colhead{$\hat{\beta}$\tablenotemark{c}}
    & \colhead{$\hat{\sigma}$\tablenotemark{d}}
    & \colhead{$\hat{\beta}$}
    & \colhead{$\hat{\sigma}$}
    & \colhead{$\hat{\beta}$}
    & \colhead{$\hat{\sigma}$}
    & \colhead{$\hat{\beta}$}
    & \colhead{$\hat{\sigma}$}
  }
  \startdata
  0.5 &  25 & $0.357^{+0.242}_{-0.246}$ & $0.784^{+0.717}_{-0.608}$
            & $0.518^{+0.513}_{-0.349}$ & $0.687^{+0.714}_{-0.650}$
            & $0.896^{+1.127}_{-0.425}$ & $0.855^{+1.616}_{-0.757}$
            & $0.513^{+0.393}_{-0.315}$ & $0.677^{+0.663}_{-0.580}$ \\
      &  50 & $0.355^{+0.166}_{-0.164}$ & $0.801^{+0.591}_{-0.528}$
            & $0.510^{+0.306}_{-0.233}$ & $0.716^{+0.601}_{-0.540}$
            & $0.898^{+0.572}_{-0.298}$ & $0.873^{+1.042}_{-0.668}$
            & $0.506^{+0.242}_{-0.212}$ & $0.717^{+0.555}_{-0.507}$ \\
      & 100 & $0.354^{+0.117}_{-0.114}$ & $0.810^{+0.488}_{-0.447}$
            & $0.506^{+0.197}_{-0.164}$ & $0.743^{+0.494}_{-0.466}$
            & $0.895^{+0.352}_{-0.218}$ & $0.885^{+0.786}_{-0.587}$
            & $0.504^{+0.162}_{-0.149}$ & $0.732^{+0.456}_{-0.429}$ \\
  1.0 &  25 & $0.190^{+0.255}_{-0.239}$ & $0.798^{+1.047}_{-0.798}$
            & $0.442^{+2.763}_{-2.167}$ & $0.610^{+1.418}_{-0.610}$
            & $0.827^{+2.293}_{-1.687}$ & $0.727^{+2.899}_{-0.727}$
            & $0.524^{+0.907}_{-0.576}$ & $0.572^{+0.903}_{-0.564}$ \\
      &  50 & $0.191^{+0.172}_{-0.164}$ & $0.839^{+0.869}_{-0.752}$
            & $0.519^{+1.816}_{-0.707}$ & $0.643^{+1.023}_{-0.643}$
            & $0.870^{+1.195}_{-0.459}$ & $0.814^{+1.754}_{-0.814}$
            & $0.519^{+0.552}_{-0.370}$ & $0.669^{+0.745}_{-0.643}$ \\
      & 100 & $0.189^{+0.121}_{-0.116}$ & $0.862^{+0.726}_{-0.640}$
            & $0.520^{+0.913}_{-0.348}$ & $0.687^{+0.784}_{-0.687}$
            & $0.895^{+0.665}_{-0.329}$ & $0.855^{+1.246}_{-0.788}$
            & $0.502^{+0.337}_{-0.242}$ & $0.714^{+0.623}_{-0.604}$ \\
  2.0 &  25 & $0.066^{+0.243}_{-0.228}$ & $0.565^{+1.797}_{-0.565}$
            & $0.036^{+2.761}_{-2.944}$ & $0.663^{+2.544}_{-0.663}$
            & $0.443^{+3.793}_{-2.836}$ & $0.000^{+2.994}_{-0.000}$
            & $0.366^{+1.468}_{-1.395}$ & $0.381^{+1.223}_{-0.362}$ \\
      &  50 & $0.067^{+0.164}_{-0.158}$ & $0.768^{+1.525}_{-0.768}$
            & $0.116^{+2.878}_{-2.951}$ & $0.743^{+2.271}_{-0.743}$
            & $0.634^{+3.276}_{-3.027}$ & $0.258^{+2.983}_{-0.258}$
            & $0.426^{+1.055}_{-0.918}$ & $0.559^{+1.082}_{-0.529}$ \\
      & 100 & $0.065^{+0.113}_{-0.106}$ & $0.843^{+1.293}_{-0.843}$
            & $0.209^{+2.936}_{-2.962}$ & $0.743^{+1.932}_{-0.743}$
            & $0.765^{+2.492}_{-2.024}$ & $0.627^{+2.928}_{-0.627}$
            & $0.444^{+0.698}_{-0.548}$ & $0.673^{+0.921}_{-0.621}$ \\
  \enddata

  \tablecomments{The values given for $\hat{\beta},$ and
    $\hat{\sigma}$ are the median and interval containing $90\%$ of
    the estimates over the simulations. For example, when $t / \tau =
    s / \sigma = 0.5$ and $n = 25$, the median value of the OLS slope
    estimator is 0.357, and $90\%$ of the values of
    $\hat{\beta}_{OLS}$ are contained within
    $0.357^{+0.242}_{-0.246}$.}

  \tablenotetext{a}{Typical value of the measurement error magnitude
    for the simulations.}
  \tablenotetext{b}{The number of data points in the simulated data
    sets.}
  \tablenotetext{c}{The estimate of the slope, $\beta$. The true value
    is $\beta = 0.5$.}

  \tablenotetext{d}{The estimate of the dispersion in the intrinsic
    scatter, $\sigma$. The true value is $\sigma = 0.75$.}

\end{deluxetable}

\end{document}